\documentclass[12pt]{article}

\usepackage{graphicx}
\usepackage{fancybox}
\usepackage{amsmath}
\usepackage{amssymb}
\usepackage{latexsym}
\usepackage{epsfig}

\newcommand{\Om}{\Omega}

\newcommand{\al}{\alpha}
\newcommand{\ep}{\epsilon}
\newcommand{\vep}{\varepsilon}

\newcommand{\la}{\lambda}

\newcommand{\R}{\mbox{R}}
\newcommand{\tR}{\widetilde{\mbox{R}}}

\newcommand{\del}{\partial}

\newcommand{\co}{{\scriptstyle \circ}}
\newcommand{\lb}{\lbrack}
\newcommand{\rb}{\rbrack}

\newcommand{\msc}[1]{\mbox{\scriptsize #1}}
\newcommand{\dsp}{\displaystyle}

\newcommand{\bc}{\Bbb C}
\newcommand{\br}{\Bbb R}
\newcommand{\bz}{\Bbb Z}

\newcommand{\bsz}{\Bbb Z}
\newcommand{\bsr}{\Bbb R}

\newcommand{\bm}[1]{\mbox{\boldmath ${#1}$}}
\newcommand{\sbm}[1]{\msc{\boldmath ${#1}$}}

\newcommand{\cO}{{\cal O}}
\newcommand{\cN}{{\cal N}}

\newcommand{\cD}{{\cal D}}

\newcommand{\cZ}{{\cal Z}}
\newcommand{\cR}{{\cal R}}

\newcommand{\tA}{\tilde{A}}

\newcommand{\tJ}{\tilde{J}}

\newcommand{\tj}{\tilde{j}}
\newcommand{\tm}{\tilde{m}}

\newcommand{\tell}{\tilde{\ell}}

\newcommand{\tQ}{\widetilde{Q}}

\newcommand{\ta}{\tilde{a}}
\newcommand{\tZ}{\widetilde{Z}}
\newcommand{\tcZ}{\widetilde{\cZ}}

\newcommand{\hv}{\widehat{v}}

\newcommand{\hc}{\hat{c}}

\newcommand{\Th}[2]{\Theta_{#1,#2}}
\renewcommand{\th}{{\theta}}

\newcommand{\ch}[2]{\mbox{ch}^{#1}_{#2}}

\newcommand{\hch}[2]{\widehat{\mbox{ch}}^{#1}_{#2}}

\newcommand{\chd}{\mbox{ch}_{\msc{\bf dis}}}

\newcommand{\hchd}{\widehat{\mbox{ch}}_{\msc{\bf dis}}}

\newcommand{\chic}{{\chi_{\msc{\bf con}}}}

\newcommand{\chid}{{\chi_{\msc{\bf dis}}}}

\newcommand{\hchid}{\widehat{\chi}_{\msc{\bf dis}}}

\newcommand{\bhchi}{\widehat{\bm{\chi}}}

\newcommand{\tr}{\mbox{Tr}}

\newcommand{\tpsi}{\tilde{\psi}}

\newcommand{\erf}{\mbox{Erf}}
\newcommand{\erfc}{\mbox{Erfc}}
\newcommand{\sgn}{\mbox{sgn}}

\newcommand{\nn}{\nonumber\\}

\newcommand{\dis}{\bf dis}
\newcommand{\con}{\bf con}

\newcommand{\reg}{\bf reg}

\renewcommand{\Re}{{\rm Re}}
\renewcommand{\Im}{{\rm Im}}

\newcommand{\any}{{}^{\forall}}

\renewcommand{\mod}{\, \mbox{mod} ~ }

\newcommand {\eqn}[1]{(\ref{#1})}

\makeatletter
\@addtoreset{equation}{section}
\def\theequation{\thesection.\arabic{equation}}
\makeatother

\begin{document}

\begin{titlepage}
 \
 \renewcommand{\thefootnote}{\fnsymbol{footnote}}
 \font\csc=cmcsc10 scaled\magstep1
 {\baselineskip=14pt
 \rightline{
 \vbox{\hbox{September, 2011}
       }}}

 \baselineskip=20pt
 
\begin{center}

{\bf \Large  Comments on Non-holomorphic Modular Forms 
\\
and 
\\
Non-compact Superconformal Field Theories 
} 

 \vskip 1.2cm
 
\noindent{ \large Yuji Sugawara}\footnote{\sf ysugawa@se.ritsumei.ac.jp}
\\

\medskip

 {\it Department of Physical Science, 
 College of Science and Engineering, \\ 
Ritsumeikan University,  
Shiga 525-8577, Japan}
 

\end{center}

\bigskip

\begin{abstract}

We extend our previous work \cite{ES-NH} on the non-compact $\cN=2$ $SCFT_2$
defined as the supersymmetric $SL(2,\br)/U(1)$-gauged WZW model.
Starting from path-integral calculations of torus partition functions 
of both the axial-type (`cigar') and the vector-type (`trumpet')  models, 
we study general models of the $\bz_M$-orbifolds and $M$-fold covers
with an arbitrary integer $M$. We then extract contributions 
of the degenerate representations (`discrete characters') 
in such a way that good modular properties are preserved.
The `modular completion' of the extended discrete characters 
introduced in \cite{ES-NH} are found to play a central role 
as suitable building blocks in every model of orbifolds or covering spaces.
We further examine a large $M$-limit (the `continuum limit'), 
which `deconstructs' the spectral flow orbits while keeping 
a suitable modular behavior. The discrete part of partition function as well as 
the elliptic genus is then expanded by the modular completions of {\em irreducible\/} 
discrete characters, which are parameterized by both continuous and discrete quantum numbers 
{\em modular transformed in a mixed way.} This limit is naturally identified 
with the universal cover of trumpet model. We finally discuss a classification of 
general modular invariants based on the modular completions of irreducible characters 
constructed above.


\end{abstract}


\setcounter{footnote}{0}
\renewcommand{\thefootnote}{\arabic{footnote}}

\end{titlepage}

\baselineskip 18pt

\vskip2cm 


\section{Introduction}

In this paper we try to extend our previous work \cite{ES-NH}
on the supersymmetric (SUSY) $SL(2,\br)/U(1)$-gauged WZW model, 
that is, the SUSY non-linear $\sigma$-model on 
2-dimensional black-hole \cite{2DBH}. 
In spite of its simplicity there are several intriguing features  
which originate from the non-compactness of target space. 
Among other things it would be surprising enough that 
this model could lead to a {\em non-holomorphic\/} 
elliptic genus \cite{Troost,ES-NH,AshokTroost}, in other words, 
{\em would-be lack of holomophic factorization} 
in the torus partition function \cite{ES-NH}.  
In more detail the torus partition function of this model 
is found to be 
expressed in the form as \cite{ES-NH}\footnote
  {Previous studies closely related to this subject have been given {\em e.g.} in
   \cite{HPT,IPT,ES-BH,IKPT,ES-C}.}
\begin{equation}
Z(\tau)= Z_{\dis}(\tau) + Z_{\con}(\tau). 
\label{Z decomp}
\end{equation}
The `continuous part' $Z_{\con}(\tau)$ is mainly contributed from free strings propagating in the asymptotic region of 
2D black-hole background, and is written in a holomorphically factorized form composed of characters of 
non-degenerate representations. 
On the other hand,  the `discrete part' $Z_{\dis}(\tau)$, which 
includes strings localized near the tip of 2D black-hole, 
would not be written in a holomorphically factorized form. 
It is modular invariant and formally expressible in an analogous way 
to rational conformal field theories (RCFTs);  
$$
Z_{\dis}(\tau)= \sum_{j, \tj} \, N_{j,\tj}\, \widehat{\chi}_j(\tau) \, \widehat{\chi}_{\tj}(\tau)^*. 
$$
However, the building blocks $\widehat{\chi}_j(\tau)$ 
are {\em no longer\/} holomorphic. They are written such as 
\begin{equation}
\widehat{\chi}_j(\tau) = \chi_j(\tau)
+ [\mbox{subleading term, function of $\tau_2$}], ~~~ (\tau_2\equiv \Im\, \tau),
\label{hchi general}
\end{equation}
where $\chi_j(\tau)$ denotes the `extended discrete character' defined by spectral flow orbits of irreducible characters generated by BPS states \cite{ES-L,ES-BH}. 
Although $\widehat{\chi}_j(\tau)$ shows the same IR behavior (around $\tau_2  \sim  + \infty$) as 
$\chi_j(\tau)$, it is never expressible in terms only of 
characters of superconformal algebra due to the $\tau_2$-dependence 
in the subleading term. 
We emphasize that, while the discrete characters $\chi_j(\tau)$ themselves are not, 
the functions $\widehat{\chi}_j(\tau)$ show simple modular behaviors 
mimicking RCFTs and are closed under modular transformations.  
When performing the S-transformation of $\chi_j(\tau)$,
a continuous term of `Mordell integral' \cite{Mordell,Watson} emerges in the similar manner to the $\cN=4$ characters \cite{ET}.
However, the subleading term exactly cancels it out, simplifying considerably the modular transformation formulas of 
$\widehat{\chi}_j(\tau)$.
Therefore, we shall call them the `modular completions' of discrete characters.


Related mathematical studies of non-holomorphic modular and Jacobi forms \cite{EZ}
seem to have been  still in new area. (See {\em e.g.} \cite{Zwegers,BOno}.)
Very roughly speaking, one finds correspondences such as
\begin{eqnarray*}
&& \mbox{discrete character}~ \longleftrightarrow ~ \mbox{mock modular form (mock theta function)},
\\
&& \mbox{modular completion} ~ \longleftrightarrow ~ \mbox{(harmonic) Maass form}.
\end{eqnarray*}
Another possible application of the theory of mock modular forms to superconformal field theories has been presented in \cite{EH}.


~

What we would like to clarify in this paper is addressed as follows;
\begin{description}
\item [(1) General $\bz_M$-orbifolds with arbitrary $M$ : ]

~

For the parafermion theory $SU(2)/U(1)$ \cite{ZF,GQ}, which is a `compact analogue' of $SL(2)/U(1)$-theory, 
general modular invariants have been classified and interpreted as some orbifolds \cite{GQ}. 
Inspired by this fact, we will examine general $\bz_M$-orbifolds of SUSY $SL(2)/U(1)$-coset 
with arbitrary $M$. This would be a natural extension of the analysis given in our previous work \cite{ES-NH}.
We especially would like to clarify the roles played by modular completions introduced in \cite{ES-NH}
in general models of orbifolds.


\item [(2) Modular completion of {\em irreducible\/} discrete character :]

~

In \cite{ES-NH} we only introduced 
the modular completions of the extended discrete characters. 
It may be a natural question what is the modular completions of the {\em irreducible\/} discrete characters.
We also would like to clarify the model including these new completions as natural building blocks.

\end{description}

~

This paper is organized as follows; 

In section 2, we demonstrate the path-integral evaluations 
of torus partition functions of both axial-type (`cigar') 
and vector-type (`trumpet') SUSY $SL(2,\br)/U(1)$ models, 
with an IR-regularization preserving good modular behaviors.

In section 3, we study general models of the $\bz_M$-orbifolds 
of cigar and the $M$-fold covers of trumpet 
with an arbitrary integer $M$, related with each other 
by the T-duality as expected \cite{Tduality-2DBH}. 
We then extract contributions of the degenerate representations 
(`discrete characters'), which are captured by 
the modular completions of extended discrete characters.
Especially, the discrete parts of partition functions of general 
orbifolds are defined so that the modular invariance is preserved. 
The `twisted' discrete partition functions are also introduced 
and they show the modular covariance. 
We further discuss a `continuum limit' by suitably taking $M\,\rightarrow \, \infty$, 
which leads us to the modular completion of {\em irreducible\/}
discrete characters. 

In section 4, we study the elliptic genera of relevant models. It will turn out that 
they are rewritten as linear combinations of 
the modular completions 
in all the cases.

In section 5, we discuss general forms of modular invariants 
when assuming the modular completions to be fundamental building blocks. 
Fourier transforms of the irreducible modular completions play a crucial role, 
and we will see that all the discrete partition functions and elliptic genera 
presented above are reexpressed in a unified way based on them.

We will summarize the main results  and give some discussions in section 6.


~


\section{Variants of $SL(2)/U(1)$ SUSY Coset Conformal Field Theories}

\subsection{Preliminaries : SUSY Gauged WZW Actions}

We shall first introduce the model which we study in this paper, summarizing relevant notations.
We consider the $SL(2,\br)/U(1)$ SUSY gauged WZW model with level $k(>0)$\footnote
   {$k$ is the level of the total $SL(2,\br)$-current including fermionic degrees of freedom, 
whose bosonic part has the level $\kappa \equiv k+2$.
}, 
which is quite familiar  \cite{KS} to have 
$\cN=2$ superconformal symmetry with central charge;
%
%
\begin{equation}
\hat{c} \equiv \frac{c}{3} =1+\frac{2}{k} ,
~~~ k\equiv \kappa-2.
\label{def c k}
\end{equation}
We restrict ourselves to cases with rational level 
$k = N/K$
($N, K \in \bz_{>0}$) for the time being,
and will later discuss the models with general levels allowed to be irrational. 
Note here that we do {\em not\,} necessarily assume that $N$ and $K$ are co-prime. 


The world-sheet action of relevant SUSY gauged WZW model in the present convention is written as
\begin{eqnarray}
 S(g,A,\psi^{\pm}, \tpsi^{\pm}) &: =& \kappa S_{\msc{gWZW}}(g,A)
 + S_{\psi}(\psi^{\pm}, \tpsi^{\pm}, A), 
\label{total action}\\
\kappa S_{\msc{gWZW}} (g,A) &: =& \kappa S^{SL(2,\bsr)}_{\msc{WZW}} (g)
+ \frac{\kappa}{\pi}\int_{\Sigma}d^2v\, 
\left\{ A_{\bar{v}} \tr \left(\frac{\sigma_2}{2} \partial_v g g^{-1}\right)
\pm
\tr \left(\frac{\sigma_2}{2}g^{-1}\partial_{\bar{v}}g\right)A_v  
\right. \nn
&& \hspace{3cm} \left. 
\pm \tr\left(\frac{\sigma_2}{2}g\frac{\sigma_2}{2} g^{-1} \right)
   A_{\bar{v}}A_v 
+ \frac{1}{2} A_{\bar{v}}A_v  \right\} , 
\label{gWZW action} \\
S^{SL(2,\bsr)}_{\msc{WZW}} (g) &: =& -\frac{1}{8\pi} \int_{\Sigma} d^2v\,
\tr \left(\partial_{\al}g^{-1}\partial_{\al}g\right) +
\frac{i}{12\pi} \int_B \,\tr\left((g^{-1}dg)^3\right) ,
\label{SL(2) WZW action} 
\\
 S_{\psi}(\psi^{\pm}, \tpsi^{\pm}, A)&: =& 
\frac{1}{2\pi}\int d^2v\, \left\{
\psi^+(\partial_{\bar{v}}+A_{\bar{v}}) \psi^-
+\psi^-(\partial_{\bar{v}}-A_{\bar{v}}) \psi^+ \right. \nn 
&& \hspace{3cm} \left. +\tpsi^+(\partial_{v}\pm A_{v}) \tpsi^-
+\tpsi^-(\partial_{v}\mp A_{v}) \tpsi^+
\right\} .
\label{fermion action}
\end{eqnarray}
In \eqn{gWZW action} and \eqn{fermion action}, the $+$ sign/ $-$ sign  is chosen for the axial-like/vector-like  
gauged WZW model, which we shall denote as 
$S^{(A)}_{\msc{gWZW}}$/$S^{(V)}_{\msc{gWZW}}$ \,
($S^{(A)}_{\psi}$/$S^{(V)}_{\psi}$ )
from now on. 
The $U(1)$ chiral gauge transformation is defined 
by 
\begin{eqnarray}
&& g~ \longmapsto ~  \Om_L \, g \,  \Om_R^{\ep} , 
\nn
&&
A_{\bar{v}}~ \longmapsto ~ 
A_{\bar{v}} - \Om_L^{-1} \del_{\bar{v}} \Om_L, ~~~
A_{v}~ \longmapsto ~ 
A_{v}- \Om_R^{-1} \del_{v} \Om_R,
\nn
&&  
\psi^{\pm} ~ \longmapsto ~ \Om_L^{\pm 1} \, \psi^{\pm},
~~~ 
\tpsi^{\pm} ~ \longmapsto ~ \Om_R^{\pm \ep}  \, \tpsi^{\pm},
\nn
&&
\hspace{4cm}
\Om_L(v,\bar{v}), ~ \Om_R(v,\bar{v}) \in e^{i\br \sigma_2}, 
\label{chiral gauge trsf}
\end{eqnarray}
where we set $\ep\equiv +1, -1$ for the axial, vector model, respectively.
The gauged WZW action $S^{(A)}_{\msc{gWZW}}$ /  $S^{(V)}_{\msc{gWZW}}$ 
is invariant under the axial/vector type gauge transformations that 
correspond to $\Om_L=\Om_R 
$ in \eqn{chiral gauge trsf}.
Both of the classical fermion actions $S^{(A)}_{\psi}$, $S^{(V)}_{\psi}$ 
\eqn{fermion action} are invariant under general chiral gauge transformations
$\Om_L$, $\Om_R$, and we assume the absence of chiral anomalies 
when $\Om_L=\Om_R$ holds.

It is well-known that this model describes the string theory on 2D Euclidean 
black-hole \cite{2DBH}. The axial-type corresponds to the cigar geometry, while 
the vector-type does to the `trumpet', which  is T-dual to the cigar 
\cite{Tduality-2DBH}. 
We will later elaborate their precise relation from the viewpoints of 
torus partition functions. 


It will be convenient to introduce alternative notations of gauged WZW actions; 
\begin{eqnarray}
&& S_{\msc{gWZW}}^{(A)} (g, h_L, h_R) :=  S^{SL(2,\bsr)}_{\msc{WZW}}(h_L g h_R) 
- S^{SL(2,\bsr)}_{\msc{WZW}}(h_L h_R^{-1}),
\label{gWZW A}
\\
&& S_{\msc{gWZW}}^{(V)} (g, h_L, h_R) :=  S^{SL(2,\bsr)}_{\msc{WZW}}(h_L g h_R) 
- S^{SL(2,\bsr)}_{\msc{WZW}}(h_L h_R).
\label{gWZW V}
\end{eqnarray}
They are indeed equivalent with \eqn{gWZW action} under the identification of 
gauge field;
\begin{equation}
A_{\bar{v}}\frac{\sigma_2}{2} = \partial_{\bar{v}} h_L h_L^{-1}, 
\hspace{1cm}
A_v \frac{\sigma_2}{2} = \ep \, \partial_v h_R h_R^{-1},
\label{identification A} 
\end{equation} 
where we set $\ep =+1  , \, (-1)$ for the axial (vector) model as before,
as one can  confirm by using the Polyakov-Wiegmann identity;
\begin{equation}
S^{SL(2,\bsr)}_{\msc{WZW}}(gh)= S^{SL(2,\bsr)}_{\msc{WZW}}(g)
+ S^{SL(2,\bsr)}_{\msc{WZW}}(h)
+ \frac{1}{\pi} \int_{\Sigma} d^2v\, \tr \left(g^{-1}\partial_{\bar{v}} g \, 
\partial_v h h^{-1}\right).
\label{PW}
\end{equation}

~


\subsection{Axial Coset : Euclidean Cigar}

We shall first focus on the axial model. We are interested in the torus partition function.
We define the world-sheet torus $\Sigma$ 
by the identifications  
$(w,\bar{w}) \sim (w+2\pi,\bar{w}+2\pi)
\sim(w+2\pi \tau, \bar{w}+2\pi \bar{\tau})$ ($\tau\equiv
\tau_1+i\tau_2$, $\tau_2>0$, 
and use the convention $v=e^{iw}$, $\bar{v}=e^{-i\bar{w}}$).
We call the cycles defined by these two identifications
as the $\al$ and $\beta$-cycles as usual.

Detailed calculations of the torus partition function have been carried out in \cite{ES-BH,ES-NH} 
based on the Wick rotated model ({\em i.e.} $H^3_+/\br_A$ supercoset, 
with $H^3_+ \equiv SL(2,\bc)/SU(2)$).
Especially, the partition function of $\tR$-sector ($\R$-sector with $(-1)^F$ insertion) 
with the $\cN=2$ moduli $z$, $\bar{z}$ ({\em i.e. } the insertion of  $e^{2\pi i \left(z J_0 - \bar{z} \tJ_0\right)}$,
where $J$, $\tJ$ are $\cN=2$ $U(1)$-currents) 
has been presented in our previous work \cite{ES-NH}. 
We shall just sketch  it here.

In the Wick rotated model $H^3_+/\br_A$, the gauge field
$ \dsp
A \equiv \left(A_{\bar{v}}d\bar{v}+ A_v dv\right)\frac{\sigma_2}{2}
$
should be regarded as a hermitian 1-form. 
Following the familiar treatment of gauged WZW models 
(see {\em e.g.} \cite{KarS,GawK,Schnitzer}), we decompose the gauge field 
as follows;
\begin{equation}
A[u]_{\bar{w}}= \partial_{\bar{w}} X  + i \partial_{\bar{w}} Y 
- \frac{u}{2\tau_2}, 
~~~  A[u]_{w}=  \partial_{w} X  - i \partial_{w} Y
- \frac{\bar{u}}{2\tau_2}
\label{Au} 
\end{equation}
where $X$, $Y$ are real scalar fields 
parameterizing the chiral gauge transformations (in the Wick rotated model);
\begin{equation}
\Om_L = e^{(X+iY) \frac{\sigma_2}{2}}, ~~~ 
\Om_R \left( \equiv \Om_L^{\dag}\right) 
= e^{(X-iY) \frac{\sigma_2}{2}},
\label{Om X Y}
\end{equation}
and $u\equiv s_1 \tau+ s_2 
$, 
$(0\leq s_1,s_2 <1)$ is the modulus of gauge field. 
To emphasize the modulus dependence of gauge field
we took the notation `$A[u]$'.
Note that the modulus parameter $u$ is normalized
so that it correctly couples with the zero-modes of $U(1)$-currents
$J^3$, $\tJ^3$ 
which should be gauged;
\begin{equation}
\left. 
- \frac{\partial}{\partial u} S(g, a[u], \psi^{\pm}, \tpsi^{\pm}) \right|_{u=0} 
= 2\pi i J^3_0, ~~~ 
\left. 
- \frac{\partial}{\partial \bar{u}} S(g,a[u], \psi^{\pm}, \tpsi^{\pm}) \right|_{u=0} 
= - 2\pi i \tJ^3_0,
\label{coupling u current}
\end{equation}
where we set 
$
\dsp 
a[u] \equiv \left(a[u]_{\bar{w}} d \bar{w} + a[u]_w dw \right) \frac{\sigma_2}{2}
\equiv 
\left(- \frac{u}{2\tau_2} d\bar{w} - \frac{\bar{u}}{2\tau_2} dw\right) 
\frac{\sigma_2}{2}.
$
For later convenience, we introduce the `monodromy function';
\begin{eqnarray}
\Phi[u] (w,\bar{w})= \frac{i}{2\tau_2}\left\{
(\bar{w}\tau-w\bar{\tau})s_1+(\bar{w}-w)s_2\right\}
\equiv \frac{1}{\tau_2} \Im (w \bar{u}),
\label{Phi u}
\end{eqnarray} 
satisfying the twisted boundary conditions;
\begin{eqnarray}
\hspace{-1cm}
\Phi[u](w+2\pi,\bar{w}+2\pi)=\Phi[u](w,\bar{w}) - 2\pi s_1,~~~
\Phi[u](w+2\pi\tau,\bar{w}+2\pi\bar{\tau})=\Phi[u](w,\bar{w}) + 2\pi s_2.
\label{bc Phi u}
\end{eqnarray}
We also introduce the notation;
\begin{equation}
h_u \equiv e^{i \Phi[u](w,\bar{w}) \frac{\sigma_2}{2}}.
\label{hu}
\end{equation}
Then, the modulus part of gauge field is expressed as 
\begin{eqnarray}
&& a[u]_{\bar{w}} = i \partial_{\bar{w}} \Phi[u] 
,
~~~ a[u]_{w} =  - i \partial_{w} \Phi[u] 
.
\label{au}
\end{eqnarray}


Including the `angle parameter' $z$ which couples with the $U(1)_R$-symmetry in  $\cN=2$ superconfomral symmetry,
the torus partition function is written as 
\begin{eqnarray}
&& Z(\tau,  z) 
= \int_{\Sigma}\frac{d^2u}{\tau_2} \, 
\int \cD\lb g, A[u], \psi^{\pm}, \tpsi^{\pm}\rb\, 
\nn
&& \hspace{2cm} \times 
\exp \left[-\kappa S^{(A)}_{\msc{gWZW}}\left(g,A[u+\frac{2}{k}z]\right) - 
S^{(A)}_{\psi}\left(\psi^{\pm},\tpsi^{\pm}, A[u+\frac{k+2}{k} z ]\right)\right],
\label{part fn A 0}
\end{eqnarray}
where $ \frac{d^2u}{\tau_2} \equiv ds_1 ds_2$ is the modular invariant measure 
of modulus parameter $u$, and we work in  the $\tR$-sector for world-sheet fermions. 
%
We can explicitly evaluate this path-integration 
by separating
the degrees of freedom of chiral gauge transformations (real scalar fields $X$ and $Y$) 
according to the standard quantization 
of gauged WZW models \cite{KarS,GawK,Schnitzer}, which renders this model 
`almost' a free conformal system. 
Namely, interactions among each sector are caused 
only through the integration of modulus $u$.  
One can easily confirm that the complex parameter $z$ precisely corresponds to the insertion of an operator 
$e^{2\pi i \left(z J_0 - \bar{z} \tJ_0\right)}$, where $J$ and $\tJ$ are 
the $\cN=2$ $U(1)$-currents in the Kazama-Suzuki model \cite{KS}. (See \cite{ES-NH} for more detail.)

To proceed further we have to path-integrate the compact boson $Y$, while the
non-compact boson $X$ is decoupled as a gauge volume. 
By using the definitions of \eqn{gWZW A}, \eqn{gWZW V} and a suitable change of integration variables, 
we obtain
\begin{eqnarray}
 Z(\tau,  z) 
&=&  \int_{\Sigma}\frac{d^2u}{\tau_2} \, 
\int \cD\lb g, Y , \psi^{\pm}, \tpsi^{\pm}, b, \widetilde{b}, c, \widetilde{c} \rb\, 
\nn
&&  \times 
\exp \left[-\kappa S^{(V)}_{\msc{gWZW}}
\left(g,h^{u+\frac{2}{k}z}, 
\left(h^{u+\frac{2}{k}z}\right)^{\dag} \right) 
+
\kappa S^{(A)}_{\msc{gWZW}}
\left(e^{iY \sigma_2},
h^{u+\frac{2}{k}z}, 
h^{u+\frac{2}{k}z}
\right)
\right]
\nn
&& \times
\exp \left[
 -2 S^{(A)}_{\msc{gWZW}}
\left(e^{iY \sigma_2},h^{u+\frac{k+2}{k}z}, 
h^{u+\frac{k+2}{k}z}\right)
- S^{(A)}_{\psi}\left(\psi^{\pm},\tpsi^{\pm}, a[u+\frac{k+2}{k} z ]\right)
\right]
\nn
&& \times 
\exp \left[
- S_{\msc{gh}}(b,\widetilde{b}, c,  \widetilde{c})\right],
\label{part fn A 1}
\end{eqnarray}
where the ghost variables have been introduced to rewrite the Jacobian factor. 
It is most non-trivial to evaluate the path-integration of the compact boson $Y$.
Its world-sheet action is evaluated as  
\begin{eqnarray}
S^{(A)}_Y(Y,u, z)& \equiv & 
- \kappa S^{(A)}_{\msc{gWZW}}
\left(e^{iY \sigma_2},
h^{u+\frac{2}{k}z}, 
\left(h^{u+\frac{2}{k}z}\right)^{\dag}
\right)
+ 2 S^{(A)}_{\msc{gWZW}}
\left(e^{iY \sigma_2},h^{u+\frac{k+2}{k}z}, 
\left(h^{u+\frac{k+2}{k}z}\right)^{\dag}\right)
\nn
& = & \frac{k}{\pi} \int_{\Sigma} d^2 v\, \del_{\bar{w}} Y^u \del_w Y^u 
- \frac{2\pi}{\tau_2} \hc \left|z\right|^2,
\label{S Y A}
\end{eqnarray}
where we set $Y^u\equiv Y + \Phi[u]$. 
Note that the linear couplings between the $U(1)$-currents $i\del_w Y$, $i\del_{\bar{w}} Y$
and the $\cN=2$ moduli $z$, $\bar{z}$ are precisely canceled out\footnote
  {Of course, this cancellation is expected  by construction 
of the $\cN=2$ superconformal algebras in the Kazama-Suzuki supercoset.}.
Since $Y^u$ satisfies the following boundary condition;
\begin{eqnarray}
&&Y^u(w+2\pi,\bar{w}+2\pi)=Y^u(w,\bar{w})- 2\pi (m_1+s_1), \nn
&&Y^u(w+2\pi\tau,\bar{w}+2\pi\bar{\tau})=Y^u(w,\bar{w})+ 2\pi (m_2+s_2), 
~~~(m_1, m_2 \in \bz), 
\label{Yu bc}
\end{eqnarray}
the zero-mode integral yields the summation over winding sectors weighted by 
the factor $e^{-\frac{\pi k}{\tau_2} \left|m_1\tau+m_2+ u\right|^2}
\equiv e^{-\frac{\pi k}{\tau_2} \left|(m_1+s_1)\tau+(m_2+s_2)\right|^2}$ determined 
by the `instanton action'.
After all, we achieve the next formula of partition function;
\begin{eqnarray}
Z(\tau,z) &=& 
\cN \, e^{\frac{2\pi}{\tau_2} \left( \hc \left|z\right|^2 - \frac{k+4}{k} z_2^2\right)}\,
\sum_{m_1,m_2\in \bz}\, \int_{\Sigma} \frac{d^2 u}{\tau_2} \, 
\left|
\frac{\th_1\left(\tau,u+\frac{k+2}{k}z\right)}{\th_1\left(\tau, u+\frac{2}{k}z\right)}
\right|^2 \,
e^{-4\pi \frac{u_2z_2}{\tau_2}} \, e^{-\frac{\pi k}{\tau_2} \left|m_1\tau+m_2+ u\right|^2}
\nn 
&\equiv & \cN \, e^{\frac{2\pi}{\tau_2} \left( \hc \left|z\right|^2 - \frac{k+4}{k} z_2^2\right)}\,
\int_{\bc} \frac{d^2 u}{\tau_2} \, 
\left|
\frac{\th_1\left(\tau,u+\frac{k+2}{k}z\right)}{\th_1\left(\tau, u+\frac{2}{k}z\right)}
\right|^2 \,
e^{-4\pi \frac{u_2z_2}{\tau_2}} \, e^{-\frac{\pi k}{\tau_2} \left|u\right|^2},
\label{part fn A}
\end{eqnarray}
where $\cN$ is a normalization constant. 
This is identified as the Euclidean cigar model whose asymptotic circle has the radius $\sqrt{\al' k}$.

To be more precise, one should make a suitable 
regularization of \eqn{part fn A},  since it shows  
an IR divergence that originates from the non-compactness of target space.
In other words, the integral of modulus $u$ 
logarithmically diverges
due to the quadratic behavior of integrand 
$\sim 1/\left|u + \frac{2}{k}z \right|^2$ near 
the point $u= - \frac{2}{k}z \in \Sigma$. 
According to \cite{ES-NH}, we take the regularization 
such that the integration region of modulus $u$ is 
replaced with 
\begin{equation}
 \Sigma(z, \ep) \equiv \Sigma \,  \setminus \, 
\left\{u= s_1 \tau + s_2 ~~;~~
-\frac{\ep}{2} - \frac{2}{k} \zeta_1 < s_1 < 
\frac{\ep}{2} - \frac{2}{k} \zeta_1,   ~ ~ 0< s_2 < 1  \right\},
\label{Sigma ep}
\end{equation}
where we set $z\equiv \zeta_1 \tau + \zeta_2$, 
$\zeta_1, \zeta_2 \in \br$, and $\ep(>0)$ denotes 
the regularization parameter. 
Then, the regularized partition function is defined as 
\begin{eqnarray}
Z_{\reg}(\tau,z; \ep) &=& 
\cN \, e^{\frac{2\pi}{\tau_2} \left( \hc \left|z\right|^2 - \frac{k+4}{k} z_2^2\right)}\,
\sum_{m_1,m_2\in \bz}\, \int_{\Sigma(z,\ep)} \frac{d^2 u}{\tau_2} \, 
\left|
\frac{\th_1\left(\tau,u+\frac{k+2}{k}z\right)}{\th_1\left(\tau, u+\frac{2}{k}z\right)}
\right|^2 \,
e^{-4\pi \frac{u_2z_2}{\tau_2}} \, e^{-\frac{\pi k}{\tau_2} \left|m_1\tau+m_2+ u\right|^2}.
\nn
&&
\label{part fn A reg}
\end{eqnarray}

One of the main results of \cite{ES-NH} is the `character decomposition' of \eqn{part fn A reg}. 
Namely, it has been shown that the partition function can be uniquely decomposed in such a form as 
\begin{eqnarray}
Z_{\reg}(\tau,z; \ep) = [\mbox{sesquilinear form of $\hchid(\tau,z)$}] +  [\mbox{sesquilinear form of $\chic(\tau,z)$}],
\label{part fn A decomp 0}
\end{eqnarray}
where $\hchid(\tau,z)$ denotes the `modular completion' of extended discrete character \cite{ES-NH}, 
while $\chic(\tau,z)$ does the extended continuous character \cite{ES-L,ES-BH} only attached with a real `Liouville momentum'  $p\in \br$ 
(above the `mass gap', in other words).
Their precise definitions and relevant formulas are summarized in Appendix C.

~

Important points are addressed as follows;
\begin{itemize}
\item The modular completion $\hchid(\tau,z)$ is non-holomorphic with respect to modulus $\tau$, but 
possesses simple modular properties: the S-transformation is closed by themselves, whereas the extended discrete character 
$\chid(\tau,z)$ is not. 
\item The second term in \eqn{part fn A decomp 0} shows a logarithmic divergence under the $\ep\,\rightarrow \, +0$ limit, which 
corresponds to the contribution from strings freely propagating in asymptotic region. 
On the other hand, the first term remains finite under $\ep\, \rightarrow \, +0$, and 
we denote it as $Z_{\dis}(\tau,z)$ (the `discrete part' of partition function). $Z_{\dis}(\tau,z)$
is modular invariant by itself, as we will later elaborate on it.   

\item It is worth pointing out that the decomposition \eqn{part fn A decomp 0} itself
{\em uniquely} determines the functional form of modular completion $\hchid (\tau,z)$. 
In fact, the subleading terms in \eqn{hchi general} are unambiguously determined by 
making `completion of the square' for terms including the extended discrete characters $\chid(\tau,z)$
\eqn{chid} in the decomposition  of $Z_{\reg}$.

\end{itemize}

~


\subsection{Vector Coset : Euclidean Trumpet}

The partition function of the vector-type model is defined in the same way as 
\eqn{part fn A 0}, with the vector-type 
gauged WZW action $S^{(V)}_{\msc{gWZW}} (g,\tA)$.
In the Wick-rotated model, we should again regard as $g \in H^3_+$,
while the gauge field $\tA$ is now parameterized as 
\begin{equation}
\tA[u]_{\bar{w}}= \partial_{\bar{w}} X  + i \partial_{\bar{w}} Y 
- \frac{u}{2\tau_2}, 
~~~  \tA[u]_{w}=  \partial_{w} X  - i \partial_{w} Y
+ \frac{\bar{u}}{2\tau_2}.
\label{tAu} 
\end{equation}
Again, the non-compact direction $X$ is anomaly free 
(vector-like),  and the compact-direction $Y$ 
is anomalous (axial-like). 
We should note that the gauge field $\tA$ is neither a hermitian 
nor an anti-hermitian 1-form. This fact originates from  
the sign difference of modulus $\bar{u}$ compared with \eqn{Au}, 
which has been chosen so that it leads to 
the same coupling to current zero-modes
as given in \eqn{coupling u current}.

Now, the wanted partition function is written as 
\begin{eqnarray}
&& \tZ(\tau,  z) 
= \int_{\Sigma}\frac{d^2u}{\tau_2} \, 
\int \cD\lb g, \tA[u], \psi^{\pm}, \tpsi^{\pm}\rb\, 
\nn
&& \hspace{2cm} \times 
\exp \left[-\kappa S^{(V)}_{\msc{gWZW}}\left(g,\tA[u+\frac{2}{k}z]\right) - 
S^{(V)}_{\psi}\left(\psi^{\pm},\tpsi^{\pm}, \tA[u+\frac{k+2}{k} z ]\right)\right].
\label{part fn V 0}
\end{eqnarray}
We can again evaluate it by using the formulas \eqn{gWZW A} and \eqn{gWZW V} as follows\footnote
   {A caution:  after separating chiral gauge transformations, the fermion action 
should get $S^{(A)}_{\psi}\left(\psi^{\pm},\tpsi^{\pm}, a[*]\right)$, rather than 
$S^{(V)}_{\psi}\left(\psi^{\pm},\tpsi^{\pm}, a[*]\right)$. It is due to our parameterization of 
gauge field $\tA$. (Recall how \eqn{tAu} includes the modulus $\bar{u}$.) This fact 
leads us to the correct fermion factor $\left|\th_1\left(\tau, u+\frac{k+2}{k} \right)\right|^2$
in the partition function \eqn{part fn V}. 
 };
\begin{eqnarray}
 \tZ(\tau,  z) 
&=&  \int_{\Sigma}\frac{d^2u}{\tau_2} \, 
\int \cD\lb g, Y , \psi^{\pm}, \tpsi^{\pm}, b, \widetilde{b}, c, \widetilde{c} \rb\, 
\nn
&&  \times 
\exp \left[-\kappa S^{(V)}_{\msc{gWZW}}
\left(g,h^{u+\frac{2}{k}z}, 
\left(h^{u+\frac{2}{k}z}\right)^{\dag} \right) 
+
\kappa S^{(V)}_{\msc{gWZW}}
\left(e^{iY \sigma_2},
h^{u+\frac{2}{k}z}, 
\left(h^{u+\frac{2}{k}z}\right)^{\dag}
\right)
\right]
\nn
&& \times
\exp \left[
 -2 S^{(V)}_{\msc{gWZW}}
\left(e^{iY \sigma_2},h^{u+\frac{k+2}{k}z}, 
\left(h^{u+\frac{k+2}{k}z}\right)^{\dag}\right)
- S^{(A)}_{\psi}\left(\psi^{\pm},\tpsi^{\pm}, a[u+\frac{k+2}{k} z ]\right)
\right]
\nn
&& \times 
\exp \left[
- S_{\msc{gh}}(b,\widetilde{b}, c,  \widetilde{c})\right].
\label{part fn V 1}
\end{eqnarray}
In deriving \eqn{part fn V 1}, we assumed that the path-integral measure 
of fermions is anomalous along the axial direction ($Y$) as opposed to 
the axial model \eqn{part fn A 1}.

The world-sheet action of compact boson $Y$ 
is now evaluated  as 
\begin{eqnarray}
S^{(V)}_Y(Y,u)& \equiv & - \kappa S^{(V)}_{\msc{gWZW}}
\left(e^{iY \sigma_2},
h^{u+\frac{2}{k}z}, 
\left(h^{u+\frac{2}{k}z}\right)^{\dag}
\right)
+ 2 S^{(V)}_{\msc{gWZW}}
\left(e^{iY \sigma_2},h^{u+\frac{k+2}{k}z}, 
\left(h^{u+\frac{k+2}{k}z}\right)^{\dag}\right)
\nn
& = & \frac{k}{\pi} \int_{\Sigma} d^2 v\, \del_{\bar{w}} Y \del_w Y 
- \frac{ik}{2\pi} \int_{\Sigma} d\Phi[u] \wedge d Y.
\label{S Y V}
\end{eqnarray}
Note that the $z$-dependence is completely canceled out 
contrary to the axial case \eqn{S Y A}. 
Moreover, the absence of quadratic term of modulus $u$ is characteristic 
for the vector-type  model. 
The second term in \eqn{S Y V} is non-dynamical and contributes to the path-integral 
just through `winding numbers';
$$
\int_{\al} d\Phi[u] = -2\pi s_1, ~~~ \int_{\beta} d\Phi[u] = -2\pi s_2, ~~~
\int_{\al} dY = 2\pi n_1, ~~~ \int_{\beta} dY = 2\pi n_2, ~~ (n_1,n_2 \in \bz). 
$$
In this way we obtain 
\begin{equation}
Z_Y^{(V)}(\tau,u) = \frac{\sqrt{k}}{\sqrt{\tau_2} \left|\eta(\tau) \right|^2}\, 
\sum_{n_1,n_2 \in \bz} \, e^{-\frac{\pi k}{\tau_2} \left|n_1 \tau + n_2 \right|^2}\,
e^{-2\pi i k (s_1 n_2 - s_2 n_1) }.
\label{Z Y V}
\end{equation}
However, we face a subtlety since $k$ is fractional in general. We recall 
$k= N/K$, and  assume that $N$ and $K$ are coprime from now on. 
The periodicity of moduli parameters $s_i\, \rightarrow \, s_i + 1$, $(i=1,2)$
would be violated unless $n_1, n_2 \in K \bz$. 
In other words, one should impose this restriction of winding numbers
to assure the consistency of functional integration.


Combining all the pieces and by taking the regularization: $\Sigma\, \rightarrow \, \Sigma(\ep, z)$
\eqn{Sigma ep}, we finally achieve the following expression 
for the vector-type coset;
\begin{eqnarray}
\hspace{-5mm}
\tZ_{\reg}(\tau,z;\ep) &=& 
\cN \, e^{-\frac{2\pi}{\tau_2} \frac{k+4}{k} z_2^2}\,
\sum_{n_1,n_2\in \bz}\, \int_{\Sigma(z,\ep)} \frac{d^2 u}{\tau_2} \, 
\left|
\frac{\th_1\left(\tau,u+\frac{k+2}{k}z\right)}{\th_1\left(\tau, u+\frac{2}{k}z\right)}
\right|^2 \,
\nn
&& \hspace{3cm} \times 
e^{-4\pi \frac{u_2 z_2}{\tau_2}} 
e^{-\frac{\pi NK}{\tau_2} \left|n_1\tau+n_2\right|^2} 
 e^{2\pi i N (n_1 s_2- n_2 s_1)}.
\label{part fn V}
\end{eqnarray}

One can also make the character decomposition for \eqn{part fn V}. 
We will work on this subject in the next section. 
Before that,  let us  first discuss aspects of the $\bz_M$-orbifolds of 
$SL(2)/U(1)$ cosets with an arbitrary integer $M$ systematically.

~


\section{General $\bz_M$-Orbifold of $SL(2)/U(1)$}

We next consider the general $\bz_M$-orbifold of $SL(2)/U(1)$-model.
We again assume a model of rational level; $k= N/K$ ($N, K \in \bz_{>0}$) 
and let $M$ be an arbitrary  divisor of $N$, setting
$N= ML$, $L \in \bz_{>0}$\footnote
   {Here we do {\em not\/} assume that $N$ and $K$ are coprime integers. 
Therefore, in case  $M$ is not a divisor of $N$, one may just replace 
$N$, $K$ with $N'\equiv NM$, $K' \equiv KM$, and all the following arguments 
are applicable. 
}.

~


\subsection{$\bz_M$-Orbifold and $M$-fold Cover}

We start with the axial model. 
Since the twisted boson $Y^u$ introduced in \eqn{S Y A} 
 is  associated  with the (asymptotic) angle coordinate of cigar geometry, 
one may consistently define the $\bz_M$-orbifold by introducing fractional winding sectors $m_1, m_2 \in \frac{1}{M}\bz$, 
leading to the torus partition function;
\begin{eqnarray}
&&Z_{\reg}^{(M)}(\tau,z; \ep) = 
\frac{1}{M} \, \sum_{\al_1,\al_2 \in \bz_M}\, Z_{\reg}^{(M)}(\tau,z\,| \al_1, \al_2; \ep) ,
\label{part fn A Z_M}
\\
&& Z^{(M)}_{\reg}(\tau,z\,| \al_1, \al_2 ;\ep) := 
k \, 
e^{- \frac{2\pi}{\tau_2}   \frac{k+4}{k} z_2^2}\,
\sum_{m_1,m_2\in \bz} \,  
\int_{\Sigma(z,\ep)} \frac{d^2 u}{\tau_2} \, 
\nn
&& \hspace{2.5cm} \times 
\left|
\frac{\th_1\left(\tau,u+\frac{k+2}{k}z\right)}{\th_1\left(\tau, u+\frac{2}{k}z\right)}
\right|^2 \,
e^{-4\pi \frac{u_2z_2}{\tau_2}} \, e^{-\frac{\pi k}{\tau_2} \left|u+ \left(m_1 + \frac{\al_1}{M}\right) \tau
+ \left(m_2 + \frac{\al_2}{M}\right)\right|^2}.
\label{Z al}
\end{eqnarray}
Here we again took the regularization \eqn{part fn A reg} and 
chose the normalization constant as $\cN=k$ so that 
$$
\lim_{\ep\,\rightarrow\, +0}\, \lim_{z\, \rightarrow \, 0}\, 
Z^{(M)}_{\reg}(\tau,z\,| 0, 0 ;\ep) = 1.
$$
We also included an modular invariant factor 
$e^{-2\pi \frac{\hc}{\tau_2} |z|^2}$ by hand to avoid unessential complexity of equations below\footnote
   {In convention adopted in this paper, the axial coset includes the factor 
$e^{2\pi \frac{\hc}{\tau_2} z_1^2}$, while the vector coset does the different factor
$e^{- 2\pi \frac{\hc}{\tau_2} z_2^2}\left(\equiv e^{2\pi \frac{\hc}{\tau_2} z_1^2}
\cdot e^{-2\pi \frac{\hc}{\tau_2} |z|^2} \right)$. We shall unify these `anomaly factors' 
to the latter one just for  convenience. Otherwise, one would be bothered about 
factors such as $e^{-2\pi \frac{\hc}{\tau_2} |z|^2}$ {\em e.g.} in \eqn{Fourier trsf 1}.
}.

The twisted partition function \eqn{Z al} behaves `almost' modular covariantly;
\begin{eqnarray}
&& Z^{(M)}_{\reg}(\tau+1,z \, | \,  \al_1,\al_2; \ep) 
= Z^{(M)}_{\reg} (\tau,z \, | \, \al_1 , \al_1+ \al_2;\ep), 
\nn
&& 
Z^{(M)}_{\reg} \left( \left. -\frac{1}{\tau}, \frac{z}{\tau}  
\, \right| \, \al_1, \al_2  ;\ep \right)=
Z^{(M)}_{\reg} (\tau,z \, | \, \al_2, -\al_1;\ep) + \cO(\ep \log \ep),
\label{modular Z al}
\end{eqnarray}
as is directly checked by the definition \eqn{Z al}. 
Note that the violation of $S$-covariance in \eqn{modular Z al} is at most 
at the order of $\cO(\ep \log \ep)$.


It is convenient to introduce the `Fourier transform' of $\eqn{Z al}$
by the next relations;
\begin{eqnarray}
\tZ^{(M)}_{\reg}(\tau,z\,| \, \beta_1, \beta_2;\ep) 
&=&  
\frac{1}{M} \,
\sum_{\al_1, \al_2 \in \bz_M}\, 
e^{2\pi i \frac{1}{M} \left(\al_1 \beta_2 - \al_2 \beta_1 \right)}\,
Z^{(M)}_{\reg} (\tau,z\,|\, \al_1, \al_2;\ep),
\label{Fourier trsf 1}
\end{eqnarray}
Using the identity 
\begin{eqnarray}
 && \hspace{-1cm}
\sum_{m_1,m_2\in \bsz}
e^{-\frac{\pi \al}{\tau_2}\left|(m_1+s_1)\tau + (m_2+s_2)\right|^2}
  e^{2\pi i (m_1 t_2-m_2 t_1)}
= \frac{1}{\al} \sum_{n_1, n_2 \in \bsz}
e^{-\frac{\pi}{\al \tau_2}\left|(n_1+t_1)\tau + (n_2+t_2)\right|^2}
  e^{2\pi i \left\lb (n_1+t_1) s_2- (n_2+t_2) s_1\right\rb}, 
\nn
&&
\hspace{10cm} 
( \Re \, \al>0, ~~~
s_i, t_i \in \br), 
\label{dualizing}
\end{eqnarray}
which is proven by the Poisson resummation formula, 
we obtain the explicit form of \eqn{Fourier trsf 1}
as 
\begin{eqnarray}
\hspace{-1cm}
\tZ^{(M)}_{\reg}(\tau,z \, | \,  \beta_1,\beta_2;\ep) &=& 
M  e^{- \frac{2\pi}{\tau_2} \frac{k+4}{k} z_2^2}\,
\sum_{n_1,n_2 \in \bz}\,
\int_{\Sigma(z,\ep)} \frac{d^2u}{\tau_2}\, 
 e^{- 4\pi \frac{u_2 z_2}{\tau_2}}\,
\left|\frac{\th_1\left(\tau, u + \frac{k+2}{k}z\right)}
{\th_1\left(\tau, u + \frac{2}{k} z \right)} \right|^2 
\nn
&& \hspace{2cm} \times 
 e^{-\frac{\pi}{k \tau_2}\left| (M n_1 +\beta_1) \tau+ (M n_2 +\beta_2) \right|^2}
\, e^{2\pi i \left\{  (M n_2 + \beta_2) s_1 - (M n_1 +\beta_1 ) s_2  \right\}}.
\label{tZ beta}
\end{eqnarray}
It shows the same  modular properties;
\begin{eqnarray}
&& \tZ^{(M)}_{\reg}(\tau+1,z \, | \,  \beta_1,\beta_2; \ep) 
= \tZ^{(M)}_{\reg} (\tau,z \, | \, \beta_1 , \beta_1+ \beta_2;\ep), 
\nn
&& 
\tZ^{(M)}_{\reg} \left( \left. -\frac{1}{\tau}, \frac{z}{\tau}  
\, \right| \, \beta_1, \beta_2  ;\ep \right)
=\tZ^{(M)}_{\reg} (\tau,z \, | \, \beta_2, -\beta_1;\ep) + \cO(\ep \log \ep).
\label{modular tZ beta}
\end{eqnarray}

~



We here present some considerations about physical interpretations of partition functions;

~

\noindent
{\bf 1.} 
From the definition \eqn{Z al} itself, it is obvious that 
$Z_{\reg}^{(M)}(\tau,z\,| \, 0,0 ;\ep) = Z_{\reg}(\tau,z;\ep) $ \eqn{part fn A reg}
for an arbitrary $M$ (up to the factor $e^{-2\pi \frac{\hc}{\tau_2}|z|^2}$).  
This fact is not surprising,  since $Z_{\reg}^{(M)}(\tau,z\,| \, 0,0 ;\ep)$ 
is associated to the untwisted sector of $\bz_M$-orbifold. 
It actually depends only on $k= N/K$, and is independent of the choice of 
pair $N$, $K$.
As already mentioned, this is identified as the Euclidean cigar model with the asymptotic radius $\sqrt{\al' k}$.
We also note that the cigar partition function \eqn{part fn A reg} is rewritten 
in the `T-dualized' form 
(with the factor $e^{-2\pi \frac{\hc}{\tau_2} |z|^2}$ included);
\begin{eqnarray}
 \hspace{-5mm}
Z_{\msc{cigar},\, \reg}(\tau,z ;\ep) \left(\equiv Z_{\reg}^{(M)}(\tau,z\,| \, 0,0 ;\ep)\right)
&=&
e^{- \frac{2\pi}{\tau_2}  \frac{k+4}{k} z_2^2}\,
\sum_{n_1,n_2 \in \bz}\,
\int_{\Sigma(z,\ep)} \frac{d^2u}{\tau_2}\, 
 e^{- 4\pi \frac{u_2 z_2}{\tau_2}}\,
\nn
&& \hspace{-2cm} \times   
\left|\frac{\th_1\left(\tau, u + \frac{k+2}{k}z\right)}
{\th_1\left(\tau, u + \frac{2}{k} z \right)} \right|^2 
\,  e^{-\frac{\pi}{k \tau_2}\left| n_1 \tau+ n_2 \right|^2}
\, e^{2\pi i \left( n_2 s_1 - n_1 s_2  \right)}, 
\label{Z trumpet}
\end{eqnarray} 
by using the identity \eqn{dualizing}. 
We identify the R.H.S of \eqn{Z trumpet} with the partition function of Euclidean trumpet 
whose asymptotic circle has the radius $\sqrt{\frac{\al'}{k}}$ \cite{Tduality-2DBH}.
In summary, we conclude 
\begin{equation}
Z_{\reg}^{(M)}(\tau,z\,| \, 0,0 ;\ep) ~ \Longleftrightarrow ~ \mbox{cigar}
~ \stackrel{\msc{T-dual}}{\Longleftrightarrow} ~ \mbox{trumpet}
, ~~~ (\mbox{indep. of } ~ M ).
\label{ZM00}
\end{equation}

~


\noindent
{\bf 2.}
The relation \eqn{Fourier trsf 1} tells us that $\tZ_{\reg}^{(M)}(\tau,z\,| \, 0,0 ;\ep)$
is identified with the $\bz_M$-orbifold of cigar.
On the other hand, the expression 
\eqn{tZ beta} is naturally interpreted as the `$M$-fold cover of trumpet', since 
the winding numbers are restricted to multiples of $M$ if compared with \eqn{Z trumpet}. 
Namely, one can summarize that 
\begin{eqnarray}
\tZ_{\reg}^{(M)}(\tau,z\,| \, 0,0 ;\ep) & = & 
\frac{1}{M}\, 
\sum_{\al_1,\al_2 \in \bz_M}\,
 Z_{\reg}^{(M)}(\tau,z\,| \, \al_1,\al_2 ;\ep)
\nn
&\Longleftrightarrow & ~  [\mbox{$\bz_M$-orbifold of cigar}]
\nn
&\stackrel{\msc{T-dual}}{\Longleftrightarrow} &  ~ [\mbox{$M$-fold cover of trumpet }].
\label{tZM00}
\end{eqnarray}
By comparing \eqn{part fn V} with \eqn{tZ beta}, we also find that 
the vector-type coset is identified with $\tZ^{(N)}_{\reg}(\tau,z \, | \,  0,0;\ep)$ 
{\em when $N$ and $K$ are coprime,} that is, 
\begin{eqnarray}
[\mbox{vector-type} ~ SL(2)_{k=N/K}/U(1) 
] &\Longleftrightarrow &  ~  [\mbox{$\bz_N$-orbifold of cigar}]
\nn
&\stackrel{\msc{T-dual}}{\Longleftrightarrow} &  ~ [\mbox{$N$-fold cover of trumpet }].
\label{interpretation vector SL(2)/U(1)}
\end{eqnarray}
In other words, we also find the equivalence;
\begin{eqnarray}
&&
[\mbox{vector-type} ~ SL(2)_{k=N/K}/U(1) 
] 
\nn
&& \hspace{2.5cm}
 ~ \Longleftrightarrow  ~  [ \bz_N\mbox{-orbifold of axial-type}
 ~ SL(2)_{k=N/K}/U(1) 
] .
\label{rel vector axial}
\end{eqnarray}
It is an obvious analogue of the familiar equivalence in parafermion theory - $SU(2)/U(1)$-coset
(see {\em e.g.} \cite{GQ,MMS});
\begin{eqnarray}
&&
\hspace{-5mm}
[\mbox{vector-type} ~ SU(2)_N/U(1) ] 
 ~ \Longleftrightarrow  ~  [ \bz_N\mbox{-orbifold of axial-type} ~ SU(2)_N/U(1) ].
\label{rel vector axial PF}
\end{eqnarray}

~


\noindent
{\bf 3.} 
In the above, we described the $\bz_M$-orbifolds of cigar and the $M$-fold covers of trumpet for an arbitrary integer $M$, 
which are T-dual with each other. 
One might then ask;  {\em how about the $M$-fold covers of cigar?}   However, they are not well-defined.
A geometrically manifest reason is the fact that 
$\pi_1(\mbox{cigar}) =0$ holds, whereas $\pi_1(\mbox{trumpet}) = \bz$.  In other words, 
one cannot restrict $m_1,m_2 \in M\bz$, $(M>1)$ in \eqn{part fn A reg}, without spoiling 
the expected periodicity of moduli $s_i\, \rightarrow \, s_i+ n_i$, $(\any n_i \in \bz)$.

In the similar sense the vector type model \eqn{part fn V} only allows $\bz_M$-orbifolds with a divisor $M$ of $N$,
while the $M$-fold covers are well-defined for arbitrarily large $M$. This is again because of  
the compatibility with periodicity of $s_i$.

~


\subsection{Discrete Parts of Partition Functions}

Let us focus on the discrete parts of various partition functions introduced above. 
It is found that all the (twisted) partition functions $Z^{(M)}_{\reg}$, $\tZ^{(M)}_{\reg}$
can be decomposed into the forms like \eqn{part fn A decomp 0}, and their discrete parts 
are uniquely determined as sesquilinear forms of the modular completion $\hchid$ \eqn{hchid}. 
Relevant analyses are quite reminiscent of those given in \cite{ES-NH} and we shall not  
detail them here.

For notational simplicity, we here introduce a new symbol 
of the modular completion;
\begin{eqnarray}
&& 
\bhchi(v,m;\tau,z) \equiv \hchid(v,a;\tau,z),  ~ ~  \mbox{with} ~ 
m \equiv v+2Ka \in \bz_{2NK}, ~~~ v=0, 1 \ldots, N-1,
\nn
&& \bhchi(v,m;\tau,z) \equiv 0, ~~ \mbox{if} ~ m-v \not\in 2K \bz.
\label{bhchi}
\end{eqnarray}
It is explicitly written as 
\begin{eqnarray}
&& 
\bhchi(v,v+2Ka;\tau,z) =  
\sum_{n\in\bz}\, 
\frac{\th_1(\tau,z)}{i\eta(\tau)^3} \, y^{2K\left(n+\frac{a}{N}\right)}
q^{NK \left(n+\frac{a}{N}\right)^2}\, 
\left\lb 
\frac{(yq^{Nn+a})^{\frac{v}{N}}}{1-yq^{Nn+a}} 
\right.
\nn
&& \hspace{1cm}
\left.
- \frac{1}{2} \sum_{r\in\bz}\, \sgn(v+Nr +0)\, 
\erfc \left(\sqrt{\frac{\pi \tau_2}{NK}} \left|v + N r\right|\right)\, 
y^{\frac{v}{N}+r} \, q^{(v+Nr)\left(n+\frac{a}{N} \right)}\,
\right],
\label{def hchid}
\end{eqnarray}
where $\erfc(x)$ denotes the error function \eqn{erfc}, 
which acts as a damping factor enough to make the power series convergent. 
See Appendix C for more detail. 
We would also use the symbol  
`$\bhchi^{[N,K]}(v,m;\tau,z) $' 
when clarifying the integer parameters $N$ and $K$\footnote
   {Note that our definition of the  extended characters 
does not only depend on the level $k\equiv N/K$, 
but on the choice of $N$ and $K$, because the sum over spectral flow 
with the flow momenta $n \in N \bz$ is taken.}.  


With this preparation the discrete part of the cigar partition function \eqn{part fn A reg} 
is written as \cite{ES-NH};
\begin{equation}
Z_{\dis}(\tau,z) = e^{-2\pi \frac{\hc}{\tau_2} z_2^2}\, 
\sum_{v=0}^{N-1} \, 
\sum_{m+\tm \in 2N\bz }\,
\bhchi(v,m ; \tau,z) \left[\bhchi(v,\tm ; \tau,z) \right]^*.
\label{Zdis} 
\end{equation}
Making similar manipulations,  we reach
the discrete partition functions for the orbifolds 
\eqn{Z al} and \eqn{tZ beta} as\footnote
   {To derive these formulas of `character decompositions',   
it seems easy to first work on $\tZ^{(M)}_{\reg}(\tau,z\,|\,\beta_1,\beta_2)$, and then 
to make use of the Fourier transformation relation \eqn{Fourier trsf 1}. In Appendix D 
we will present an explicit calculation following that given in \cite{ES-NH}.
}
\begin{eqnarray}
\hspace{-5mm}
Z_{\dis}^{(M)}(\tau,z\,|\al_1,\al_2) 
& = & e^{-2\pi \frac{\hc}{\tau_2} z_2^2}\, 
\sum_{v=0}^{N-1} \, 
\sum_{m+\tm \equiv 2L \al_1~ (\msc{mod}~ 2LM) }\,
e^{2\pi i \frac{\al_2}{2KM} (m-\tm)}\,
\bhchi(v,m ; \tau,z) \left[\bhchi(v,\tm ; \tau,z) \right]^*
\nn
&& 
\label{Zdis bhchi}
\\
\hspace{-5mm}
\tZ_{\dis}^{(M)}(\tau,z\,|\beta_1,\beta_2) 
& = & e^{-2\pi \frac{\hc}{\tau_2} z_2^2}\, 
\sum_{v=0}^{N-1} \, 
\sum_{\stackrel{m+\tm \in 2L \bz}{m - \tm \equiv 2K \beta_1~ (\msc{mod}~ 2KM)} }\,
e^{2\pi i \frac{\beta_2}{2LM} (m+\tm)}\,
\bhchi(v,m ; \tau,z) \left[\bhchi(v,\tm ; \tau,z) \right]^*.
\nn
&& 
\label{tZdis bhchi}
\end{eqnarray}
It turns out that they possess the strict modular covariance;
\begin{eqnarray}
&& \hspace{-1cm}
Z^{(M)}_{\dis}(\tau+1,z \, | \,  \al_1,\al_2) 
= Z^{(M)}_{\dis} (\tau,z \, | \, \al_1 , \al_1+ \al_2), ~~~ 
Z^{(M)}_{\dis} \left( \left. -\frac{1}{\tau}, \frac{z}{\tau}  
\, \right| \, \al_1, \al_2 \right)
=Z^{(M)}_{\dis} (\tau,z \, | \, \al_2, -\al_1) ,
\nn
&& \hspace{-1cm}
\tZ^{(M)}_{\dis}(\tau+1,z \, | \,  \beta_1,\beta_2) 
= \tZ^{(M)}_{\dis} (\tau,z \, | \, \beta_1 , \beta_1+ \beta_2), ~~~ 
\tZ^{(M)}_{\dis} \left( \left. -\frac{1}{\tau}, \frac{z}{\tau}  
\, \right| \, \beta_1, \beta_2 \right)
=\tZ^{(M)}_{\dis} (\tau,z \, | \, \beta_2, -\beta_1) .
\nn
&&
\label{modular Z dis}
\end{eqnarray}
These nice features are expected from the fact that the discrete part is uniquely 
determined from the regularized partition functions and    
the regularization parameter $\ep$ is removed safely. 
We will later  prove these formulas directly from 
the modular transformation formulas of $\bhchi(*,*;\tau,z)$ given in 
\eqn{S hchid}, \eqn{T hchid}.


~


Several remarks are in order;

~

\noindent
{\bf 1.} All the Fourier and $\bz_M$-orbifold (or $M$-fold cover) relations previously 
discussed are still preserved after separating the discrete parts. 
For instance, 
we obtain
\begin{eqnarray}
\tZ^{(M)}_{\dis}(\tau,z\,| \, \beta_1, \beta_2) 
&=&  
\frac{1}{M} \,
\sum_{\al_1, \al_2 \in \bz_M}\, 
e^{2\pi i \frac{1}{M} \left(\al_1 \beta_2 - \al_2 \beta_1 \right)}\,
Z^{(M)}_{\dis} (\tau,z\,|\, \al_1, \al_2),
\label{dis Fourier trsf 1}
\end{eqnarray}
corresponding to \eqn{Fourier trsf 1}.
This fact results from the uniqueness of the relevant decompositions, and  
is easy to check by comparing directly \eqn{Zdis bhchi} and \eqn{tZdis bhchi}.

~


\noindent
{\bf 2.}
All the discrete partition functions introduced here are independent 
of the choice of pair $(N,K)$ as long as $k=N/K$ is fixed.
As already mentioned, the same statement is obvious by definition 
for the regularized partition functions $Z^{(*)}_{\reg}$, and 
again it results from the uniqueness of decompositions.  
It is, however,  non-trivial to check it directly, 
because the modular completion $\bhchi(*,*;\tau,z)$ {\em does\/} depend on 
the choice of pair $(N,K)$, not only on $k\equiv N/K$.
In doing it, the next identity is useful;
\begin{equation}
\sum_{j\in \bz_s}\,\bhchi^{[s N, s K]}(s v, s\{v+2K(a+ Nj)\} \, ; \tau,z)
= \bhchi^{[N,K]} (v,v+2Ka ;\tau,z)
\label{identity bhchi}
\end{equation}
where we made it explicit the $(N,K)$ dependence of $\bhchi(*,*;\tau,z)$. 
This identity is proven by using the definition of $\bhchi(*,*;\tau,z)$ 
\eqn{hchid}.

We also point out that $Z^{(M)}_{\dis}(\tau,z\,|\, 0,0) $ 
does not depend even on $M$ in the similar manner to
 $Z^{(M)}_{\reg}(\tau,z\,|\, 0,0;\ep) $. 
Namely, 
$
Z^{(M)}_{\dis}(\tau,z\,|\, 0,0) = Z_{\dis}(\tau,z),
$
holds for an arbitrary $M \in \bz_{>0}$. 


~

\noindent
{\bf 3.} 
When observing  the structures of discrete partition functions \eqn{Zdis bhchi} and \eqn{tZdis bhchi}, 
one may notice a reminiscence of general modular invariants in the parafermion theory \cite{GQ}. 
It gets clearer if focusing on the one for the $\bz_M$-orbifold;
\begin{eqnarray}
\hspace{-5mm}
Z_{\dis}^{(M)}(\tau,z) &\equiv & 
\tZ_{\dis}^{(M)}(\tau,z\,|0,0) 
=  e^{-2\pi \frac{\hc}{\tau_2} z_2^2}\, 
\sum_{v=0}^{N-1} \, 
\sum_{\stackrel{m+\tm \in 2L \bz}{m - \tm \in 2KM \bz} }\,
\bhchi(v,m ; \tau,z) \left[\bhchi(v,\tm ; \tau,z) \right]^*
\nn
&=& 
e^{-2\pi \frac{\hc}{\tau_2} z_2^2}\, 
\sum_{v=0}^{N-1} \, 
\sum_{r \in \bz_{2 KM},\,  s \in \bz_{L}}\,
\bhchi(v,Lr + KM s ; \tau,z) \left[\bhchi(v, Lr-KM s ; \tau,z) \right]^*,
\label{Zdis ZM orb}
\end{eqnarray}
which is indeed modular invariant.
Especially, for models with integer levels $k=N$, $(K=1)$, 
we find\footnote
   {$Z^{(N)}_{\dis}(\tau,z)$ always has a diagonal form for an arbitrary level $k=N/K$. 
However, $Z_{\dis}(\tau,z)$ gets anti-diagonal only when taking  $K=1$.
} 
\begin{eqnarray*}
&& Z_{\dis}(\tau,z) ~ \lb \mbox{axial } SL(2)_N/U(1) \rb ~ \longleftrightarrow ~ \mbox{anti-diagonal  modular  invariant} ,
\\
&& Z^{(N)}_{\dis}(\tau,z) ~ \lb \mbox{vector } SL(2)_N/U(1) \rb~ \longleftrightarrow ~ \mbox{diagonal  modular  invariant}.
\end{eqnarray*}
Of course, this resemblance is not surprising because such a structure has its origin in  the $U(1)$-coset and taking suitable orbifolds. 
A crucial difference is that there exist infinitely many inequivalent models here, 
while having at most a finite number of modular invariants in the parafermion theory. 
Note that one can choose an arbitrarily large integer $M$ by scaling $(N,K)\, \rightarrow \, (MN, MK)$
if necessary.

~


\subsection{`Continuum Limit'}

Now, let us discuss a limit of `infinite order orbifold', which should be
also described by the universal cover of trumpet model by T-duality;
$$
[\mbox{`$\bz$-orbifold' of cigar}] \cong [\mbox{universal cover of trumpet}].
$$
We shall take the large $M$-limit with keeping $ k \equiv N/K$ fixed.
Since we require $M$ divides $N$, 
we have to make $N$ (and $K$) to be large simultaneously. 
Hence, it is simplest to set $M=N$ first. We then take the large $N$-limit with $k$ fixed. 
Since $Z^{(N)}_{\dis}(\tau,z)$ is a diagonal modular invariant as mentioned above,  
we have 
$$
\lim_{z,\bar{z} \,\rightarrow\, 0} Z^{(N)}_{\dis} (\tau,z) = N,
$$
corresponding to the $N$ Ramond vacua with quantum numbers 
$$
h=\tilde{h} = \frac{\hc}{8} , ~~~ Q=\tQ = \frac{v}{N}-\frac{1}{2}, 
~~ (v=0,1,\ldots, N-1).
$$
Hence, the naive large $N$-limit apparently diverges 
due to more and more dense distribution of Ramond vacua. 
We rather define the `continuum limit' by 
\begin{equation}
Z_{\dis}^{(\infty)}(\tau,z) := \lim_{\stackrel{N\,\rightarrow\, \infty}{k\equiv N/K }\, \msc{fixed}}\, 
\frac{1}{N} \, Z^{(N)}_{\dis} (\tau,z) ,
\label{Z infty}
\end{equation}  
so that it is contributed from the continuum of Ramond vacua 
located on the segment  
$
-\frac{1}{2} \leq Q \leq \frac{1}{2}
$ 
with unit density.
Since the extended characters are defined to be the spectral flow orbits 
with flow momenta $n\in N\bz$,  taking the large $N$-limit should deconstruct the orbits, leaving 
the irreducible characters. 
An explicit calculation yields 
\begin{eqnarray}
Z^{(\infty)}_{\dis}(\tau,z)
&=& 
e^{-2\pi \frac{\hc}{\tau_2}z_2^2} \, 
\sum_{r \in\bz} \, \frac{1}{k} \int_0^k d\la 
\,  \hchd(\la, r;\tau,z) \,\left[ \hchd(\la, r ;\tau,z)\right]^*,
\label{cont Z}
\end{eqnarray}
where $\hchd(\la, r;\tau,z)$ denotes the modular completion 
of {\em irreducible\/} discrete character \eqn{hchd} 
defined by
\begin{eqnarray}
\hspace{-5mm}
\hchd(\la, n;\tau,z) 
&: =& 
\frac{\th_1(\tau,z)}{i\eta(\tau)^3} y^{\frac{2n}{k}}q^{\frac{n^2}{k}}\, \left\lb 
\frac{(yq^n)^{\frac{\la}{k}}}{1-yq^n} 
- \frac{1}{2} \sum_{\nu \in \la + k\bz}\, \sgn(\nu+0)\, 
\erfc \left(\sqrt{\frac{\pi \tau_2}{k}} \left|\nu \right|\right)\, 
\left(yq^n\right)^{\frac{\nu}{k}}
\right]
\nn
& \equiv & \lim_{\stackrel{N\,\rightarrow\, \infty}
{k\equiv N/K \, \msc{fixed}}}\, \bhchi^{[N,K]}(v, v+2K n ; \tau,z ),
\hspace{1.5cm}
(0\leq \la \leq k, ~~ n\in \bz),
\label{def hchd}
\end{eqnarray}
with the identification $v= K\la$.
The infinite $r$-sum appearing in \eqn{cont Z} converges as long as $\tau_2 >0$,
which is shown based on the following facts;
\begin{itemize}
\item The irreducible discrete character $\chd(\la,n)$ \eqn{ch d} (the first term in the first line of \eqn{def hchd}) obviously 
well behaves under the large $|n|$-limit.

\item 
By utilizing a simple inequality
$\erfc\left(\sqrt{\frac{\pi \tau_2}{k}}|\nu| \right) 
\leq 
e^{-\pi \tau_2 \frac{\nu^2}{k}}
$, 
the difference $\dsp \hchd(\la,n) - \chd(\la,n) =: f_n(\tau,z) \frac{\th_1(z)}{i\eta(\tau)^3} $ 
is evaluated as 
$$
\left|f_n(\tau,z) \right| \leq C \left|y^{\frac{n}{k}}\right| \, e^{-\pi \tau_2 \frac{n^2}{k}},
$$
with some constant $C$ independent of $n$. 
\end{itemize}
The modular transformation formulas of \eqn{def hchd} are presented 
in \eqn{S hchd}, \eqn{T hchd}. 
It may be amusing that $\hchd(*,*;\tau,z)$ includes both continuous and discrete quantum
numbers, which are S-transformed in a mixed way. 

More generally we define the twisted partition function at this limit by
\begin{eqnarray}
\hspace{-0.5cm}
\tZ^{(\infty)}_{\dis} (\tau, z\,|\, n_1, n_2 ) &:= &  
\lim_{\stackrel{N\,\rightarrow\, \infty}{k\equiv N/K }\, \msc{fixed}}\, 
\frac{1}{N} 
\, \tZ^{(N)}_{\dis}(\tau,z\,|\,n_1,n_2)
\nn
&=& 
e^{- 2\pi \frac{\hc}{\tau_2}z_2^2} \, 
\sum_{r \in\bz} \, \frac{1}{k} \int_0^k d\la
\, e^{2\pi i \frac{n_2}{k}\left(\la + n_1 + 2r \right) } 
\,  \hchd (\la, n_1+r;\tau,z) \,\left[ \hchd (\la, r ;\tau,z)\right]^*,
\nn
&&
\hspace{9cm} (n_1,n_2 \in \bz). 
\label{cont tZ}
\end{eqnarray}
It satisfies the modular covariance relation 
\begin{eqnarray}
&& \hspace{-1cm}
\tZ^{(\infty)}_{\dis}(\tau+1,z \, | \,  n_1, n_2) 
= \tZ^{(\infty)}_{\dis} (\tau,z \, | \, n_1 , n_1+n_2), ~~~ 
\tZ^{(\infty)}_{\dis} \left( \left. -\frac{1}{\tau}, \frac{z}{\tau}  
\, \right| \, n_1, n_2 \right)
=\tZ^{(\infty)}_{\dis} (\tau,z \, | \, n_2, -n_1) ,
\nn
&&
\label{modular cont tZ}
\end{eqnarray}
which is obvious by construction.


~

We add a few remarks:

~

\noindent
{\bf 1.}
Although we took the above limit by setting $M=N$ first, 
the definition \eqn{cont tZ} does not depend on this procedure.
Namely, one can confirm that 
\begin{eqnarray}
\tZ^{(\infty)}_{\dis} (\tau, z\,|\, n_1, n_2 ) &= &  
\lim_{\stackrel{M\,\rightarrow\, \infty, \, \, M\,\msc{divides}\, N,}{ 
k\equiv N/K \, \msc{fixed}}}\, 
\frac{1}{M} 
\, \tZ^{(M)}_{\dis}(\tau,z\,|\,n_1,n_2)
\label{cont tZ def2}
\end{eqnarray}
holds.
In fact, one can reconstruct the discrete partition function 
\eqn{tZdis bhchi} from \eqn{cont tZ} as follows
$(\beta_1,\beta_2 \in \bz_M )$;
\begin{equation}
\tZ^{(M)}_{\dis} (\tau,z\,|\, \beta_1,\beta_2) 
= M \sum_{m_1,m_2\in\bz}\, \tZ^{(\infty)}_{\dis} 
(\tau,z \, |\, \beta_1+ M m_1, \beta_2+ M m_2),
\label{reconstruction tZdis}
\end{equation}
and \eqn{cont tZ def2} is readily obtained from this fact.

~


\noindent
{\bf 2.} 
The modular completion of irreducible character \eqn{def hchd}
is well-defined for general level $k\in \br_{>0}$ and depends continuously on it. 
Therefore, even though assumed the rational $k$ to derive the above results, 
one may regard the formulas \eqn{cont Z}, \eqn{cont tZ} as correct ones for general $k$.
However, the extended discrete characters and their modular completion are 
well-defined only for the rational levels.

~

%
%
\noindent
{\bf 3.}
In deriving \eqn{cont Z}, \eqn{cont tZ}, 
we first integrated  
the modulus $u\equiv s_1\tau+s_2$ out, 
and then took the large $N$-limit.  
If we instead take the large $N$-limit first,
we would obtain 
\begin{eqnarray}
Z^{(\infty)}_{\reg}(\tau,z;\ep) &=& 
k e^{-\frac{2\pi}{\tau_2}\frac{k+4}{k}z_2^2}\,
\sum_{m_1,m_2\in \bz}\,
\int_{0}^{1} dt_1 \, 
\int_{0}^{1} dt_2 \,
\int_{\Sigma(z;\ep)} \frac{d^2u}{\tau_2}\, 
\nn
&& ~~~ \times 
\left|\frac{\th_1\left(\tau, u+\frac{k+2}{k}z\right)}
{\th_1\left(\tau, u+\frac{2}{k}z\right)}\right|^2
\,e^{-4\pi \frac{u_2z_2}{\tau_2}}\, e^{-\frac{\pi k}{\tau_2}
\left|u+(t_1+m_1)\tau+(t_2+m_2)\right|^2},
\label{Zreg cont}
\end{eqnarray}
as the continuum limit of the regularized partition function. 
Each twisted sector of the orbifold is parameterized by 
the continuous parameters $t_1$, $t_2$.
Note here that 
the order of integrations is crucial. 
One has to first integrate the modulus parameter $u$
to achieve the above result \eqn{cont Z}.    
If inverting the order, we would not gain sensible discrete partition functions with expected modular properties\footnote
   {In \cite{ES-nh-P} a refined regularization scheme has been introduced, and based on it, the $u$ and $t$ integrals 
have been defined with no subtlety so as to 
commute with each other.}. 
This subtlety comes from the presence/absence of Gaussian 
factor in the $u$-integral, which is necessary  to utilize 
some contour deformation techniques that plays an important role 
to extract the discrete part. (See \cite{ES-NH} and Appendix D.)  
If making first the $t$-integration (combined with the summation over $m_i$), 
the Gaussian factor drops off, 
and one could not extract suitable discrete parts. 

~


\subsection{Direct Proof of Modular Covariance}

As promised before, let us here present a direct proof of the modular covariance 
of discrete partition functions \eqn{Zdis  bhchi}, \eqn{tZdis bhchi}, 
and \eqn{cont tZ}, that is, the formulas \eqn{modular Z dis} and  
\eqn{modular cont tZ} based on the modular transformation formulas of 
$\bhchi(*,*;\tau,z)$.  
We only focus on 
$\tZ^{(M)}_{\dis}(\tau,z \,|\, \al, \beta)$ 
\eqn{tZdis bhchi}.
After that, remaining  formulas for the other partition functions 
\eqn{Zdis bhchi} and \eqn{cont tZ} can be readily derived by using 
the Fourier relation \eqn{dis Fourier trsf 1} and the definition 
\eqn{cont tZ} itself.

For our purpose it is more convenient to rewrite \eqn{tZdis bhchi} in terms of 
the original notation of modular completion 
$\hchid (v,a;\tau,z)$,  in which the discrete parameters $v$ and $a$ are unconstrained; 
\begin{equation}
\tZ_{\dis} (\tau, z\,| \, \al, \beta) =  
e^{-2\pi \frac{\hc}{\tau_2}z_2^2} \, 
\sum_{(v,a, \ta) \in \cR (M,\al)}
\, e^{2\pi i \frac{\beta}{N} \left\{v+K(a+\ta)  \right\}} \,
\hchid (v,a;\tau,z) \left[\hchid (v, \ta ;\tau,z)\right]^*.
\label{tZdis al beta}
\end{equation}
Here the range of summation is defined by 
\begin{eqnarray}
\cR(M,\al) & :=  & \left\{
(v, a, \ta) \in \bz \times \bz_{N} \times \bz_{N}~;~
0\leq v \leq N-1, \right. 
\nn
&& \hspace{2cm} \left. 
a-\ta \equiv \al~ (\mod M), ~  v+K(a+\ta) \in L\bz
\right\}.
\label{R M al}
\end{eqnarray}

Consider first the T-transformation. 
Due to the formula \eqn{T hchid} we obtain 
\begin{eqnarray}
\tZ^{(M)}_{\dis}(\tau+1,z\,|\,\al,\beta) &=& 
e^{-2\pi \frac{\hc}{\tau_2}z_2^2} \, 
\sum_{(v,a, \ta) \in \cR (M, \al)}
\, e^{2\pi i \frac{\beta}{N} \left\{v+K (a+\ta)  \right\}} \,
e^{2\pi i \left\{
\frac{K}{N} (a^2-\ta^2) + \frac{v}{N} (a-\ta)
\right\}}
\nn
&& \hspace{2cm} \times 
\hchid (v,a;\tau,z) \, \left[\hchid (v,  \ta ;\tau,z)\right]^*
\nn
&=& e^{-2\pi \frac{\hc}{\tau_2}z_2^2} \, 
\sum_{(v,a, \ta) \in \cR (M,\al)}
\, e^{2\pi i \frac{\al+ \beta}{N} \left\{v+K(a+\ta)  \right\}} \,
\hchid (v,a;\tau,z) \, \left[\hchid (v,  \ta ;\tau,z)\right]^*
\nn
&=& \tZ^{(M)}_{\dis}(\tau,z\,|\, \al, \al+\beta).
\label{check T dis} 
\end{eqnarray}
On the other hand, 
the calculation of S-transformation is much more complicated. 
We first obtain from \eqn{S hchid}
\begin{eqnarray}
\tZ^{(M)}_{\dis} \left(-\frac{1}{\tau}, \frac{z}{\tau} \,|\, \al,\beta \right)
&=& e^{-2\pi \hc \frac{z_2^2}{\tau_2}}\,
\sum_{(v,a, \ta) \in \cR (M, \al)} \, 
e^{2\pi i \frac{\beta}{N} \left\{v+K(a+\ta)  \right\}} \,
\nn
&& \hspace{5mm}
\times \sum_{v_L', v_R'=0}^{N-1}\,
\sum_{a_L', a_R' \in \bz_{N}}\, 
\frac{1}{N} e^{-2\pi i \frac{v a_L'+ v'_L a + 2K a a_L'}{N}}\,
\frac{1}{N} e^{2\pi i \frac{v a_R'+ v'_R \ta + 2K \ta a_R'}{N}}\,
\nn
&& \hspace{3cm} 
\times \hchid (v_L',a_L';\tau,z) \, 
\left[ \hchid (v_R',  a_R' ;\tau,z)\right]^*
\nn
&=& e^{-2\pi \hc \frac{z_2^2}{\tau_2}}\,
\sum_{\hv\in L \bz_{M}} \,
\sum_{a,\ta \in \bz_N, \, a-\ta \in \al + M\bz}\, 
 \sum_{v_L', v_R'=0}^{N-1}\,
\sum_{a_L', a_R' \in \bz_{N}}\, 
\nn
&& \hspace{5mm}
\times e^{2\pi i \frac{\beta \hv}{N}} \,
\frac{1}{N} e^{-2\pi i \frac{\hv a_L'+ v'_L a + 2K a a_L'}{N}}\,
\frac{1}{N} e^{2\pi i \frac{\hv a_R'+ v'_R \ta + 2K \ta a_R'}{N}}\,
e^{2\pi i \frac{K}{N} (a+\ta) (a_L'-a_R')}
\nn
&& \hspace{3cm} 
\times \hchid (v_L',a_L';\tau,z) \, 
\left[ \hchid (v_R',  a_R' ;\tau,z)\right]^*.
\label{eq1}
\end{eqnarray}
In the second line we set $\hv := v+ K(a+\ta)$.
The $\hv$-summation is easy to carry out, yielding 
Kronecker symbol;
$$
\delta^{(M)}_{a_L'-a_R',\beta} \equiv 
\left\{
\begin{array}{ll}
1 & ~~ a_L'-a_R' \equiv \beta~ (\mod\, M) \\
0 & ~~ \mbox{otherwise,}
\end{array}
\right.
$$
Moreover, rewriting $\ta = a-\al+ Mm$, $(m\in \bz_L)$, 
we obtain
\begin{eqnarray}
\hspace{-1cm}
[\mbox{R.H.S of \eqn{eq1}}]
&=& e^{- 2\pi \hc \frac{z_2^2}{\tau_2}}\,
\sum_{a \in \bz_N}\,
\sum_{m\in \bz_L}\, 
 \sum_{v_L', v_R'=0}^{N-1}\,
\sum_{a_L', a_R' \in \bz_{N}}\, 
\nn
&& \hspace{5mm}
\times M \delta^{(M)}_{a_L'-a_R', \beta} \,
\frac{1}{N^2} \,e^{-2\pi i \frac{a}{N} \left( v_L'-v_L' \right)} \,
e^{2\pi i \frac{(-\al)}{N} \left\{v_R' + K(a_L'+a_R') \right\}} \,
e^{2\pi i \frac{m}{L} \left\{v_R' + K(a_L'+a_R') \right\} }
\nn
&& \hspace{4cm} 
\times \hchid (v_L',a_L';\tau,z) \, 
\left[ \hchid (v_R',  a_R' ;\tau,z)\right]^*
\nn
&=& e^{-2\pi \hc \frac{z_2^2}{\tau_2}}\, 
\sum_{v_L', v_R'=0}^{N-1}\,
\sum_{a_L', a_R' \in \bz_{N}}\, 
 \delta^{(M)}_{a_L'-a_R', \beta} \,
\delta^{(N)}_{v_L',v_R'} \,
\delta^{(L)}_{v_R'+K(a_L'+a_R'),0}
\nn
&& \hspace{2.5cm}
\times
e^{2\pi i \frac{(-\al)}{N} \left\{v_R' + K(a_L'+a_R') \right\}} \,
\hchid (v_L',a_L';\tau,z) \, 
\left[ \hchid (v_R',  a_R' ;\tau,z)\right]^*
\nn
&=& e^{-2\pi \hc \frac{z_2^2}{\tau_2}}\, 
\sum_{(v', a_L', a_R') \in \cR(M,\beta)}\, 
e^{2\pi i \frac{(-\al)}{N} \left\{v_R' + K(a_L'+a_R') \right\}} \,
\hchid (v',a_L';\tau,z) \, 
\left[ \hchid (v', a_R' ;\tau,z)\right]^*
\nn
&\equiv& 
\tZ_{\dis}^{(M)}(\tau,z\,|\, \beta,-\al).
\label{check S dis}
\end{eqnarray}
This is the desired result.


~


\section{Elliptic Genera}


The elliptic genus \cite{Witten-E} is a nice tool to examine 
important aspects of any $\cN=2$ superconformal filed theory.
It is defined by formally setting $\bar{z}=0$ in the
partition function of $\tR$-sector, while leaving $z$ at a generic value. 
In \cite{ES-NH} we studied the elliptic genera of the cigar $SL(2,\br)/U(1)$ theory
\eqn{part fn A} and  some orbifolds of it. 
These analyses were based on the character decomposition of partition function mentioned above
and also the direct evaluation of path-integration.
For instance, the elliptic genus of the simplest cigar model is written (in the notation adopted here) as 
\begin{eqnarray}
 \cZ(\tau,z) &=& \lim_{\ep\,\rightarrow \, +0}\,  
k \, e^{\frac{\pi}{k \tau_2} z^2}\,
\sum_{m_1,m_2\in \bz} \,  
\int_{\Sigma(z,\ep)} \frac{d^2 u}{\tau_2} \, 
\frac{\th_1\left(\tau,u+\frac{k+2}{k}z\right)}
{\th_1\left(\tau, u+\frac{2}{k}z\right)}
 \,
\nn
&& \hspace{2cm} \times 
e^{2\pi i \frac{u_2}{\tau_2}z} \, 
e^{-\frac{\pi k}{\tau_2} \left|u+ 
m_1 \tau + m_2 \right|^2}
\nn
&=& \sum_{v=0}^{N-1} \, 
\sum_{\stackrel{a\in\bz_N}{v+Ka \in N\bz}}\,
\bhchi(v,v+2Ka \,; \, \tau,z) .
\label{cZ}
\end{eqnarray}
The first line is derived  by formally setting $\bar{z}=0$ in 
the regularized partition function \eqn{part fn A reg}. 
Since the integrand of \eqn{cZ} has at most simple poles, 
the $\ep \rightarrow  +0$ limit converges, and it is easy to confirm 
that the first line shows an expected modular behavior;
\begin{equation}
\cZ(\tau+1,z) = \cZ(\tau,z),
\hspace{1cm}
\cZ\left(- \frac{1}{\tau}, \frac{z}{\tau}\right) 
= e^{i\pi \frac{\hc}{\tau}z^2}\, \cZ(\tau,z),
\label{modular cZ}
\end{equation}
which is characteristic for the Jacobi form of weight 0 and index $\hc/2$. 
We should, however, emphasize that $\cZ(\tau,z)$ is not holomorphic with respect to $\tau$ 
in a sharp contrast with compact superconformal models.

On the other hand, the second line can be derived in two ways;
\begin{description}
\item[(i)] 
One may  substitutes  the formula of Witten Indices \eqn{WI} into  
the `character decomposition' \eqn{Zdis}.
Then, it is easy to obtain the second line of \eqn{cZ}. 
Note that the `continuous part' in the decomposition \eqn{part fn A decomp 0} 
does not contribute when setting $\bar{z}=0$.


\item[(ii)]
One can also directly show the equality between the first and the second lines in 
\eqn{cZ} by means of Poisson resummation and  some contour deformation techniques. 
See \cite{ES-NH} for the detail. It would be important as a cross check of calculations. 
Furthermore, this equality implies good modular behaviors of the modular completions $\bhchi(*,*;\tau,z)$. 
In fact, it seems  easiest to derive the modular S-transformation 
formula \eqn{S hchid} based on such an identity for the orbifold cases discussed below.

\end{description}


Now, the main purpose of this section is to make generalizations of 
this analysis to various orbifold models given in the previous section. 
Again we assume $k\equiv N/K$ ($N$, $K$ are not necessarily assumed  
to be coprime) and let $N= ML$. 
Explicit analyses are almost parallel.
First of all, corresponding to \eqn{Z al}, we shall introduce the elliptic genus with the $\bz_M$-twisting as  
\begin{eqnarray}
&&
 \cZ^{(M)}(\tau,z\,| \al_1, \al_2) \equiv  \lim_{\ep\,\rightarrow \, +0}\,  
k \, e^{\frac{\pi}{k \tau_2} z^2}\,
\sum_{m_1,m_2\in \bz} \,  
\int_{\Sigma(z,\ep)} \frac{d^2 u}{\tau_2} \, 
\frac{\th_1\left(\tau,u+\frac{k+2}{k}z\right)}{\th_1\left(\tau, u+\frac{2}{k}z\right)}
 \,
\nn
&& \hspace{3cm} \times 
e^{2 \pi i \frac{u_2}{\tau_2}z } \, e^{-\frac{\pi k}{\tau_2} \left|u+ \left(m_1 + \frac{\al_1}{M}\right) \tau
+ \left(m_2 + \frac{\al_2}{M}\right)\right|^2}.
\label{cZ al}
\end{eqnarray}
One can show 
\begin{equation}
\cZ^{(M)}(\tau,z\,|\, \al_1,\al_2) 
 =  \sum_{v=0}^{N-1} \, 
\sum_{\stackrel{a\in\bz_N}{v+Ka \equiv L\al_1 \, (\msc{mod} \, N)}}\,
e^{2\pi i \frac{\al_2}{M} a} \,
\bhchi(v,v+2Ka \,; \, \tau,z) ,
\label{cZ bhchi}
\end{equation}
in the same way as \eqn{cZ}. 
$\cZ^{(M)}(\tau,z\,|\, \al_1,\al_2) $ shows the modular covariance;
\begin{eqnarray}
&& \cZ^{(M)}(\tau+1,z \, | \, \al_1, \al_2) = \cZ^{(M)}(\tau,z\,|\, \al_1, \al_1+\al_2),
\nn
&& 
\cZ^{(M)}\left(- \frac{1}{\tau}, \frac{z}{\tau}\, | \, \al_1, \al_2 \right) 
= e^{i\pi \frac{\hc}{\tau}z^2}\, \cZ^{(M)}(\tau,z\, | \, \al_2, -\al_1).
\label{modular cZ al}
\end{eqnarray}
These relations are straightforwardly proven 
by using the `path-integral representation'
\eqn{cZ al}. One can also derive it based on the `character decomposition' 
\eqn{cZ bhchi} and the modular transformation formulas 
\eqn{S hchid}, \eqn{T hchid} as in \eqn{check T dis}, \eqn{check S dis}.

It is also useful to introduce the `Fourier transform' of \eqn{cZ al} 
as in \eqn{Fourier trsf 1};
\begin{eqnarray}
\tcZ^{(M)}(\tau,z\,| \, \beta_1, \beta_2) 
&=&   \frac{1}{M} \,\sum_{\al_1, \al_2 \in \bz_M}\, 
e^{2\pi i \frac{1}{M} \left(\al_1 \beta_2 - \al_2 \beta_1 \right)}\,
\cZ^{(M)} (\tau,z\,|\, \al_1, \al_2),
\label{EG Fourier trsf 1}
\end{eqnarray}
which again has the modular covariance. 
We can explicitly compute it
as 
\begin{eqnarray}
 \tcZ^{(M)}(\tau,z\,| \beta_1, \beta_2) &=& \lim_{\ep\,\rightarrow \, +0}\,  
M \, e^{\frac{\pi}{k \tau_2} z^2}\,
\sum_{n_1,n_2 \in \bz}\,
\int_{\Sigma(z,\ep)} \frac{d^2u}{\tau_2}\, 
\frac{\th_1\left(\tau, u + \frac{k+2}{k}z\right)}
{\th_1\left(\tau, u + \frac{2}{k} z \right)} 
\nn
&& \hspace{2cm} \times 
e^{2 \pi i \frac{u_2}{\tau_2}z}\,
 e^{-\frac{\pi}{k \tau_2}\left| (M n_1 +\beta_1) \tau+ (M n_2 +\beta_2) \right|^2}
\, e^{2\pi i \left\{  (M n_2 + \beta_2) s_1 - (M n_1 +\beta_1 ) s_2  \right\}}
\nn
& =&  \sum_{v=0}^{N-1} \, 
\sum_{\stackrel
{v+Ka \in L \bz}{a \equiv \beta_1 \, (\msc{mod}\, M)}}\,
e^{2\pi i \frac{\beta_2}{N} (v+Ka)} \,
\bhchi(v,v+2Ka \,; \, \tau,z) .
\label{tcZ bhchi}
\end{eqnarray}

It is obvious that 
\begin{equation}
\cZ^{(M)}(\tau,z\,|\, 0,0) = \cZ(\tau,z) , ~~~ (\mbox{for} ~ \any M) 
\end{equation}
by definition, while
we have 
\begin{equation}
\tcZ^{(M)}(\tau,z\,|\, 0, 0)= \sum_{v=0}^{N-1} \, 
\sum_{\stackrel
{v+Ka \in L \bz}{a \in M\bz}} \,
\bhchi(v,v+2Ka \,; \, \tau,z) ,
\end{equation}
which is identified with the elliptic genus of $\bz_M$-orbifold of cigar;
\begin{equation}
\cZ^{(M)}(\tau,z) \equiv \frac{1}{M} \sum_{\al_1,\al_2\in\bz_M}\,
\cZ^{(M)}(\tau,z\,|\,\al_1,\al_2).
\label{ell ZM orb}
\end{equation}
Especially, in the special case $M=N$, we obtain
\begin{equation}
\tcZ^{(N)}(\tau,z\,|\, \beta_1,\beta_2) 
 =  \sum_{v=0}^{N-1} \, 
e^{2\pi i \frac{\beta_2}{N} (v+K\beta_1)} \,
\bhchi(v,v+2K\beta_1 \,; \, \tau,z) ,
\label{tcZ bhchi Z_N}
\end{equation}
and
\begin{equation}
\cZ^{(N)} (\tau,z) \equiv 
\tcZ^{(N)}(\tau,z\,|\, 0, 0)= \sum_{v=0}^{N-1} \, 
\bhchi(v,v\,; \, \tau,z) ,
\end{equation}
which has been already  derived in \cite{Troost,ES-NH},
and nicely expressible in terms of the modular completion of 
the higher level Appell function \cite{Appell} given in \cite{Zwegers}.


Finally, for the `continuum limit' describing the universal cover of trumpet,
we achieve the next formula;
\begin{eqnarray}
\tcZ^{(\infty)}(\tau, z\, | \, n_1,n_2) 
&:= &
\lim_{\stackrel{M\,\rightarrow\, \infty,\, M\,\msc{divides}\,N }{k\equiv N/K \, \msc{fixed}}}\,
\frac{1}{M} \, \tcZ^{(M)}(\tau,z\,|\, n_1, n_2)
\nn
&=& \frac{1}{k} \int_0^k d\la
\, e^{2\pi i \frac{n_2}{k}\left(\la + n_1 \right) } 
\, \hchd (\la, n_1;\tau,z) , ~~~ (n_1,\,n_2 \in \bz).
\label{cont EG twisted}
\end{eqnarray}
The formula \eqn{cont EG twisted} reads as the Fourier transform of 
the modular completion of irreducible discrete character $\hchd(*,*;\tau,z)$.
It is worthwhile to point out that 
the modular covariance of $\tcZ^{(\infty)}(\tau, z\, | \, n_1,n_2) $ 
(in the same form as \eqn{modular cZ al}) 
is equivalent with the modular transformation formulas 
of $\hchd(*,*;\tau,z)$ \eqn{S hchd}, \eqn{T hchd}.
In section 5, we will make use of \eqn{cont EG twisted} as fundamental building blocks 
to construct general modular invariants.

We remark that the order of taking limits is again important.  
In deriving \eqn{cont EG twisted}, we have to first make the integration of modulus $u$,
and then should take the large $M$-limit. 
If inverting the order, the Gaussian factor for the $u$-integration drops off,
and the resultant elliptic genus does not coincide with \eqn{cont EG twisted}.
It does not possess  expected properties of elliptic genus 
contrary to \eqn{cont EG twisted}.
In other words, the `path-integral representation' of elliptic genera 
(such as the first line of \eqn{cZ}) would be unambiguously defined only for 
finite order orbifolds\footnote
   {Again we would like to notice that 
the continuum limit \eqn{cont EG twisted} has been suitably defined 
without any subtlety 
of the order of taking limits in \cite{ES-nh-P}. 
Furthermore, the $\la$-integral appearing in \eqn{cont EG twisted} has been explicitly carried out in that paper, yielding 
a very simple expression of \eqn{cont EG twisted} (or \eqn{Psi n}). 
See also the footnote of page 18. 
}. 

~


\section{Classification  of General Modular Invariants}

We finally discuss a classification of general modular invariants. 
In this section we consider models with arbitrary $k>0$, which may be 
irrational. 
To be more specific,  the questions we would like to ask are as follows; 
\begin{description}
\item[(1)]
What is the most general candidate of elliptic genus possessing the suitable modular property
(as a Jacobi form), when assuming the expression; 
\begin{equation}
\cZ(\tau,z) = \frac{1}{k} \int_0^k d\la\, \sum_{n\in \bz} \, \rho(\la,n) \, \hchd (\la, n;\tau,z), 
\label{cZ general 0}
\end{equation}
with some density function $\rho(\la, n)$? 

\item[(2)]
What is the most general candidate of `discrete partition function' that is modular invariant 
and written in the form as
\begin{eqnarray}
&& \hspace{-1cm}
Z_{\dis}(\tau,z) = e^{- 2\pi \frac{\hc}{\tau_2} z_2^2}\, 
\frac{1}{k^2} \int_0^k d\la_L \, \int_0^k d\la_R \,
\sum_{n_L,n_R \in \bz} 
\nn
&& 
\hspace{2cm} 
\times \sigma(\la_L,\la_R, n_L, n_R) \, \hchd (\la_L, n_L;\tau,z)
\, \left[\hchd (\la_R, n_R;\tau,z)\right]^* 
\label{Zdis general 0}
\end{eqnarray}
with some density function $\sigma(\la_L,\la_R, n_L,n_R)$? 

\end{description}


At first glance, it would not appear so easy to solve these classification problems.    
However, the problems get easier if we rephrase them  by using 
\eqn{cont EG twisted} as the fundamental building blocks.
Note that \eqn{cont EG twisted} is just the Fourier transform of 
$\hchd$ and the generality of assumption is maintained. 
For notational simplicity we introduce
\begin{equation}
\Psi_{\sbm{n}} (\tau,z) := \tcZ^{(\infty)}(\tau, z\, | \, n_1,n_2), \hspace{1cm} (\bm{n} \equiv (n_1,n_2) \in \bz^2),
\label{Psi n}
\end{equation}
where the R.H.S is given in \eqn{cont EG twisted}.
The modular property of the building block $\Psi_{\sbm{n}}$ is very simple;
\begin{eqnarray}
&& \Psi_{\sbm{n}} \left(
\frac{a\tau+b}{c\tau+d}, \frac{z}{c\tau+d}
\right)
= e^{i\pi \hc \frac{c z^2}{c\tau+d}}\,
\Psi_{\sbm{n} A} (\tau,z), 
\hspace{1cm}
\any A \equiv 
\left(
\begin{array}{cc}
a & b \\
c & d
\end{array}
\right) 
\in SL(2,\bz)
\label{modular Psi_n}
\end{eqnarray}
It is also convenient to introduce the `spectral flow operator' 
$s_{\sbm{n}}$ ($\any \bm{n} \equiv (n_1,n_2)  \in \bz^2$) 
defined by 
\begin{equation}
s_{\sbm{n}}\cdot f (\tau,z) = (-1)^{n_1+n_2} e^{2\pi i \frac{n_1n_2}{k}}
q^{\frac{\hc}{2} n_1^2} y^{\hc n_1} \, 
f(\tau, z+n_1\tau+n_2), 
\end{equation}
for any function $f(\tau,z)$. 
Then, $\Psi_{\sbm{n}}$ can be expressed as 
\begin{equation}
\Psi_{\sbm{n}}(\tau,z) = s_{\sbm{n}}\cdot \Psi_{\sbm{0}}(\tau,z) \equiv 
s_{\sbm{n}}\cdot \frac{1}{k} \int_0^k d\la\, \hchd (\la,0;\tau,z).
\end{equation}
Moreover, because of the useful relation; 
\begin{equation}
s_{\sbm{m}}\, \co \, s_{\sbm{n}} =
 e^{-2\pi i \frac{1}{k} \langle \sbm{m}, \sbm{n} \rangle }\, s_{\sbm{m+n}},
\end{equation}
where $\langle ~~, ~~ \rangle$ denotes an $SL(2,\bz)$-invariant product;
\begin{equation}
\langle \bm{m}, \bm{n} \rangle :=  m_1 n_2 - m_2 n_1, 
\hspace{1cm}  (\bm{m}\equiv (m_1,m_2), ~~ \bm{n}\equiv (n_1,n_2))
\end{equation}
we obtain
\begin{eqnarray}
&& s_{\sbm{m}} \cdot \Psi_{\sbm{n}} = 
e^{-2\pi i \frac{1}{k} \langle \sbm{m}, \sbm{n} \rangle}\,
\Psi_{\sbm{m}+\sbm{n}}, \hspace{1cm} (\any \bm{m}, \bm{n} \in \bz^2).
\label{sflow Psi_n}
\end{eqnarray}

Now, since $\Psi_{\sbm{n}} $ is the Fourier transform of $\hchd$, 
one can rewrite the ansatz \eqn{cZ general 0} in the form of
\begin{equation}
\cZ^{[a]}(\tau,z) = \sum_{\sbm{n}\in \bz^2}\, a(\bm{n})\, \Psi_{\sbm{n}} (\tau,z).
\label{cZ general}
\end{equation}
Moreover, by requiring the modular property;
\begin{equation}
 \cZ^{[a]} \left(
\frac{a\tau+b}{c\tau+d}, \frac{z}{c\tau+d}
\right)
= e^{i\pi \hc \frac{c z^2}{c\tau+d}}\, \cZ^{[a]} (\tau,z),
\label{cZa modular}
\end{equation}
and using \eqn{modular Psi_n}, 
one can obtain a simple constraint;
\begin{equation}
a(\bm{n} A) = a(\bm{n}), \hspace{1cm} (\any A \in SL(2,\bz)).
\end{equation}
%
%
If satisfying it, this is actually the most general function 
possessing the modular property \eqn{cZa modular}
as long as assuming the ansatz \eqn{cZ general 0}.


Similarly, the most general modular invariant of 
the form \eqn{Zdis general 0}
is given by 
\begin{equation}
Z^{[c]}_{\dis} (\tau,z) = e^{-2\pi \frac{\hc}{\tau_2} z_2^2}\, 
\sum_{\sbm{n}_L, \sbm{n}_R \in \bz^2}\, c(\sbm{n}_L,\sbm{n}_R)\, 
\Psi_{\sbm{n}_L} (\tau,z)\, \left[\Psi_{\sbm{n}_R} (\tau,z) \right]^*,
\label{Zdis general}
\end{equation} 
where $c(\bm{n}_L,\bm{n}_R)$ are arbitrary $SL(2,\bz)$-invariant coefficients;
\begin{equation}
c(\bm{n}_L A, \, \bm{n}_R A) = c(\bm{n}_L, \bm{n}_R), \hspace{1cm} (\any A \in SL(2,\bz)).
\nonumber
\end{equation}

~

Let us elaborate on the concrete examples:

~

\noindent
\underline{\bf 1. Elliptic Genera:}

\begin{description}
\item[(i)] $a(\bm{n}) = \delta_{\sbm{n},\sbm{0}}$ : 

This is the simplest example, and \eqn{cZ general} reduces to the elliptic genus of universal cover of the trumpet;
\begin{equation}
\cZ^{[a]}(\tau,z) = \Psi_{\sbm{0}}(\tau,z) =  \cZ^{(\infty)}(\tau,z) \equiv \frac{1}{k} \int_0^k d\la\, \hchd (\la,0;\tau,z)
\label{cZ general ex 1}
\end{equation}


\item[(ii)] $a(\bm{n}) = 1$, $(\any \bm{n}\in \bz^2)$ : 

In this case, we obtain
\begin{equation}
\cZ^{[a]}(\tau,z) 
= \sum_{\sbm{n}\in \bz^2}\, \Psi_{\sbm{n}} (\tau,z) .
\label{cZ general ex 2}
\end{equation}
In case of $k=N/K$ this is equal
the elliptic genus of original cigar model $\cZ(\tau,z)$ \eqn{cZ}.
Note that the expression \eqn{cZ general ex 2} is still well-defined 
even if $k$ is not rational, and one may regard it as the general formula 
of $\cZ(\tau,z)$.


\item[(iii)] $a(\bm{n}) = M \delta^{(M)}_{\sbm{n},\sbm{0}}$ : 

In this case \eqn{cZ general} reduces to 
\begin{equation}
\cZ^{[a]}(\tau,z) = M \sum_{\sbm{n} \in \bz^2}\, \Psi_{M\sbm{n}}(\tau,z).
\label{cZ general ex 3}
\end{equation}
In case of $k=N/K$ this is equal
the elliptic genus of $\bz_M$-orbifold of cigar $\tZ^{(M)}(\tau,z) 
\equiv \tcZ^{(M)}(\tau,z\,|\,0,0)$
 \eqn{ell ZM orb}.
More generally, we can prove
\begin{equation}
 \tcZ^{(M)}(\tau,z| \beta_1,\beta_2) =
M \sum_{\sbm{n} \in \bz^2}\, \Psi_{\sbm{\beta}+ M\sbm{n}}(\tau,z) , 
~~~ (\any \bm{\beta} \equiv (\beta_1,\beta_2) \in \bz_M),
\end{equation} 
which is an analogue of the identity \eqn{reconstruction tZdis}.
One may again regard it as a formula applicable to general $k$, not necessarily rational.

\end{description}

~


\noindent
\underline{\bf 2. Discrete Partition Functions:}

\begin{description}
\item[(i)] $c(\bm{n}_L , \bm{n}_R) = \delta_{\sbm{n}_L,\sbm{n}_R}$ : 

This has the diagonal form, and \eqn{Zdis general} corresponds to the universal cover of trumpet \eqn{cont Z};
\begin{equation}
Z_{\dis}^{[c]}(\tau,z) = e^{-2\pi \frac{\hc}{\tau_2} z_2^2}\,  \sum_{\sbm{n} \in \bz^2}\, 
\Psi_{\sbm{n}}(\tau,z) \left[\Psi_{\sbm{n}} (\tau,z)\right]^* 
=  Z_{\dis}^{(\infty)}(\tau,z) ,
\label{Zdis general ex 1}
\end{equation}
More generally the twisted partition function \eqn{cont tZ} is rewritten as ($\bm{n}\equiv (n_1,n_2) \in \bz^2$)
\begin{eqnarray}
Z_{\dis}^{(\infty)}(\tau,z\,|\, n_1,n_2) &=& e^{-2\pi \frac{\hc}{\tau_2} z_2^2}\,  \sum_{\sbm{m} \in \bz^2}\, 
s_{\sbm{n}}\cdot
\Psi_{\sbm{m}}(\tau,z) \left[\Psi_{\sbm{m}} (\tau,z)\right]^* 
\nn
&=& e^{-2\pi \frac{\hc}{\tau_2} z_2^2}\,  \sum_{\sbm{m} \in \bz^2}\, 
e^{-2\pi i \frac{1}{k} \langle \sbm{n}, \sbm{m} \rangle} \,
\Psi_{\sbm{m+n}}(\tau,z) \left[\Psi_{\sbm{m}} (\tau,z)\right]^* .
\end{eqnarray}


\item[(ii)] $c(\bm{n}_L , \bm{n}_R) = e^{-2\pi i \frac{1}{k} \langle \sbm{n}_L, \sbm{n}_R \rangle}$ : 

This coefficient is indeed $SL(2,\bz)$-invariant. The corresponding partition function is 
found to be the one of the original cigar model \eqn{Zdis}; 
\begin{equation}
Z_{\dis}^{[c]}(\tau,z) = e^{-2\pi \frac{\hc}{\tau_2} z_2^2}\,  \sum_{\sbm{n}_L,\sbm{n}_R \in \bz^2}\, 
e^{-2\pi i \frac{1}{k} \langle \sbm{n}_L, \sbm{n}_R \rangle} \,
\Psi_{\sbm{n}_L}(\tau,z) \left[\Psi_{\sbm{n}_R} (\tau,z)\right]^* 
=  Z_{\dis}(\tau,z) ,
\label{Zdis general ex 2}
\end{equation}
Again this formula works for general $k \in \br_{>0}$.


\item[(iii)] $c(\bm{n}_L,\bm{n}_R) = M e^{-2\pi i \frac{1}{k} \langle \sbm{n}_L, \sbm{n}_R \rangle} 
\delta^{(M)}_{\sbm{n}_L,\sbm{n}_R}$ : 

In this case \eqn{cZ general} corresponds to the $\bz_M$-orbifold;
\begin{equation}
Z_{\dis}^{[c]}(\tau,z) = e^{-2\pi \frac{\hc}{\tau_2} z_2^2}\, M  
\sum_{\stackrel{\sbm{n}_L,\sbm{n}_R \in \bz^2}{\sbm{n}_L-\sbm{n}_R \in M\bz^2}}\, 
e^{-2\pi i \frac{1}{k} \langle \sbm{n}_L, \sbm{n}_R \rangle} \,
\Psi_{\sbm{n}_L}(\tau,z) \left[\Psi_{\sbm{n}_R} (\tau,z)\right]^* 
=  Z^{(M)}_{\dis}(\tau,z) ,
\label{Zdis general ex 3}
\end{equation}
In fact, one can show that it is equal to \eqn{Zdis ZM orb} in case of $k=N/K$.

More generally, we can prove
\begin{eqnarray}
\tZ_{\dis}^{(M)}(\tau,z\,|\, \beta_1,\beta_2) &=& 
e^{-2\pi \frac{\hc}{\tau_2} z_2^2}\, M  \sum_{\stackrel{\sbm{n}_L,\sbm{n}_R \in \bz^2}{\sbm{n}_L-\sbm{n}_R \in \sbm{\beta} + M \bz^2}}\, 
e^{-2\pi i \frac{1}{k} \langle \sbm{n}_L, \sbm{n}_R \rangle} \,
\Psi_{\sbm{n}_L}(\tau,z) \left[\Psi_{\sbm{n}_R} (\tau,z)\right]^* ,
\nn
&& 
\hspace{6cm}  (\any \bm{\beta} \equiv (\beta_1,\beta_2) \in \bz_M),
\end{eqnarray} 
which is essentially the identity equivalent 
with \eqn{reconstruction tZdis}.

\end{description}


~


\section{Summary and Discussions}

We first summarize new results given in this paper that extend the work \cite{ES-NH};
\begin{itemize}
\item 
We have studied the $\bz_M$-orbifolds of the cigar SUSY $SL(2,\br)/U(1)$-coset with a rational level $k=N/K$
and the $M$-fold covers of trumpet $SL(2,\br)/U(1)$-coset with an arbitrary integer $M$,
which are related by the T-duality relations.
We have extracted contributions of the BPS representations (`discrete characters') 
in such a way that good modular properties are preserved.
The modular completions of the extended discrete characters introduced in \cite{ES-NH} 
work as  suitable building blocks in every model of orbifold or covering space. 
Especially, the elliptic genera are 
naturally expressed in terms only of these 
modular completions in all the models.  


\item We have further examined a large $M$-limit (the `continuum  limit'). 
Both the discrete part of partition function and the elliptic genus 
have been expanded by the modular completions of {\em irreducible\/}
discrete characters. In a sharp contrast with any RCFT, 
we have an infinite number of building blocks at this limit, 
including both discrete and continuous quantum numbers
{\em being S-transformed in a mixed way}. 
It would not be likely that modular invariants 
of such a kind have been known until now.  
This limit is geometrically identified with the universal cover 
of trumpet model.


\item We also discussed cases of general level $k$, 
{\em allowed to be non-rational.} 
The discrete part is still well-defined, and is well captured 
by the modular completions of irreducible characters. 
General modular invariants for the discrete part 
are classified for an arbitrary $k$, in which 
the Fourier transforms of irreducible modular completions \eqn{Psi n} play 
a crucial role. 
In the cases of rational $k$, general modular invariants have forms mimicking the parafermion theory.
However, there exist an infinite number of inequivalent classes of modular invariants 
as opposed to the parafermion case.  We also note that the solution \eqn{Zdis general}
includes much broader class of modular invariants. For instance, non-diagonal modular invariants  
with $\la_L\neq \la_R$ are possible, although it is not yet obvious whether they are physically interpretable 
in the context of 2D black-hole models.   

\end{itemize}

~

Probably, one of important issues we should discuss is
the origin of non-holomorphicity of  elliptic genus, 
or equivalently, an apparent lack of holomorphic factorization
in the relevant models.
If respecting simple modular properties, 
which is our stand point in this paper, 
the modular completions should be fundamental building blocks. 
On the other hand, if respecting the holomorphic factorization, 
the partition functions are expanded by an infinite number of 
extended or irreducible continuous (non-BPS) characters 
{\em which may be non-unitary\/},
in addition to the discrete (BPS) ones.
In the latter picture, 
the emergence of non-unitary characters would cause 
an IR-instability. However, summing up infinite 
characters and making an analytic continuation, we 
can gain an IR-stable partition function, as is consistent 
with the former picture.
This power series would promote extra singularities, which cancel
the zeros of $\th_1$-factors and contribute to the Witten index. 
This  is indeed the possible origin of non-holomorphicity of the elliptic genus. 
More detailed study on this issue 
should be one of our future works.

Such an incompatibility between the simple modular behavior
and the holomorphic factorization seems to be a characteristic 
feature of non-compact conformal field theories that include 
both discrete and continuous spectra of normalizable 
states. Searching other models showing similar aspects 
is surely an interesting subject.

~


\section*{Acknowledgments}

The author would like to thank T. Eguchi for valuable discussions.
This work is  supported in part  by the Grant-in-Aid
for Scientific Research (C) No.23540322 from Japan Society for the
Promotion of Science (JSPS).


~

~



\section*{Appendix A: ~ Conventions for Theta Functions}

\setcounter{equation}{0}
\def\theequation{A.\arabic{equation}}

We assume $\tau\equiv \tau_1+i\tau_2$, $\tau_2>0$ and 
 set $q:= e^{2\pi i \tau}$, $y:=e^{2\pi i z}$;
 \begin{equation}
 \begin{array}{l}
 \dsp \th_1(\tau,z)=i\sum_{n=-\infty}^{\infty}(-1)^n q^{(n-1/2)^2/2} y^{n-1/2}
  \equiv 2 \sin(\pi z)q^{1/8}\prod_{m=1}^{\infty}
    (1-q^m)(1-yq^m)(1-y^{-1}q^m), \\
 \dsp \th_2(\tau,z)=\sum_{n=-\infty}^{\infty} q^{(n-1/2)^2/2} y^{n-1/2}
  \equiv 2 \cos(\pi z)q^{1/8}\prod_{m=1}^{\infty}
    (1-q^m)(1+yq^m)(1+y^{-1}q^m), \\
 \dsp \th_3(\tau,z)=\sum_{n=-\infty}^{\infty} q^{n^2/2} y^{n}
  \equiv \prod_{m=1}^{\infty}
    (1-q^m)(1+yq^{m-1/2})(1+y^{-1}q^{m-1/2}), \\
 \dsp \th_4(\tau,z)=\sum_{n=-\infty}^{\infty}(-1)^n q^{n^2/2} y^{n}
  \equiv \prod_{m=1}^{\infty}
    (1-q^m)(1-yq^{m-1/2})(1-y^{-1}q^{m-1/2}) .
 \end{array}
\label{th}
 \end{equation}
 \begin{eqnarray}
 \Th{m}{k}(\tau,z)&=&\sum_{n=-\infty}^{\infty}
 q^{k(n+\frac{m}{2k})^2}y^{k(n+\frac{m}{2k})} .
 \end{eqnarray}
 We also set
 \begin{equation}
 \eta(\tau)=q^{1/24}\prod_{n=1}^{\infty}(1-q^n).
 \end{equation}
%
%
The spectral flow properties of theta functions are summarized 
as follows;
\begin{eqnarray}
 && \th_1(\tau, z+m\tau+n) = (-1)^{m+n} 
q^{-\frac{m^2}{2}} y^{-m} \th_1(\tau,z) ~, \nn
&& \th_2(\tau, z+m\tau+n) = (-1)^{n} 
q^{-\frac{m^2}{2}} y^{-m} \th_2(\tau,z) ~, \nn
&& \th_3(\tau, z+m\tau+n) = 
q^{-\frac{m^2}{2}} y^{-m} \th_3(\tau,z) ~, \nn
&& \th_4(\tau, z+m\tau+n) = (-1)^{m} 
q^{-\frac{m^2}{2}} y^{-m} \th_4(\tau,z) ~, \nn
&& \Th{a}{k}(\tau, 2(z+m\tau+n)) = 
q^{-k m^2} y^{-2 k m} \Th{a+2km}{k}(\tau,2z)~.
\label{sflow theta}
\end{eqnarray}

~

\section*{Appendix B:~ Relevant Path-integral Formulas}

\setcounter{equation}{0}
\def\theequation{B.\arabic{equation}}



We summarize several path-integral formulas relevant to our analysis.
\begin{description}

\item[\underline{$H^3_+$-sector:} \cite{Gaw}]
\begin{eqnarray}
&& Z^{(V)}_g(\tau,u) \equiv  \int \cD g\, \exp\left[-\kappa S^{(V)}(g,\, h^u,\, h^{u\, \dag})\right]
=  \frac{e^{2\pi \frac{u_2^2}{\tau_2}}}{\sqrt{\tau_2}
|\th_1(\tau,u)|^2}, \label{Gaw formula 1} \\ 
&& Z^{(A)}_g(\tau,z) \equiv
\int \cD g\, \exp\left[-\kappa S^{(A)}(g,\, h^u,\, h^{u\, \dag})\right] 
=   \frac{e^{2\pi \frac{u_2^2}{\tau_2} 
- \pi \kappa \frac{|u|^2}{\tau_2}}}{\sqrt{\tau_2}
|\th_1(\tau,u)|^2}. \label{Gaw formula 2} 
\end{eqnarray}

\item[\underline{$U(1)$-sector ($Y$-sector)}:]
\begin{eqnarray}
\hspace{-1cm}
Z^{(A)}_Y(\tau,u)&\equiv  & 
\int \cD Y \, \exp\left[-\frac{1}{\pi \al'} \int d^2w\, 
\left|\partial_{\bar{w}} Y^u\right|^2 \right] \nn
&=&  
\frac{\sqrt{k}}{\sqrt{\tau_2}\left|\eta(\tau)\right|^2}\,
\sum_{m_1,m_2\in \bsz}\,
e^{-\frac{\pi k}{\tau_2}
\left|(m_1+s_1)\tau+(m_2+s_2)\right|^2}
\label{Z Y A formula}
\\
\hspace{-1cm}
Z^{(V)}_Y(\tau,u)&\equiv  & 
\int \cD Y \, \exp\left[-\frac{1}{\pi \al'} \int d^2w\, 
\left|\partial_{\bar{w}} Y \right|^2
- \frac{ik}{2\pi} \int_{\Sigma} d\Phi[u] \wedge d Y\right] \nn
&=&  \frac{\sqrt{k}}{\sqrt{\tau_2} \left|\eta (\tau) \right|^2}\, 
\sum_{\stackrel{n_1,n_2 \in  \bz}{kn_i \in \bz}} \, e^{-\frac{\pi k}{\tau_2} \left|n_1 \tau + n_2 \right|^2}\,
e^{-2\pi i k (s_1 n_2 - s_2 n_1) }.
\label{Z Y V formula}
\end{eqnarray}
In the axial case \eqn{Z Y A formula}, we set $Y^u \equiv Y + \Phi[u] $,
which satisfies the twisted boundary condition \eqn{Yu bc}.
In the vector case \eqn{Z Y V formula}, we assume $k$ to be rational.


\item[\underline{fermion sector:}]
\begin{eqnarray}
Z_{\psi}\left(\tau,u\right)
&\equiv & 
\int \cD\lb\psi^{\pm},\tpsi^{\pm}\rb\,
\exp\left[ -S^{(A)}_{\psi}
\left(\psi^{\pm},\tpsi^{\pm},a[u]\right)\right] 
=
e^{-2\pi \frac{u_2^2}{\tau_2}}
\, \left|\frac{\th_1(\tau,u)}
{\eta(\tau)}\right|^2.
\label{Z psi} 
\end{eqnarray}


\item[\underline{ghost sector:}]
\begin{eqnarray}
Z_{\msc{gh}}(\tau)
&\equiv  & 
\int \cD\lb b, \tilde{b}, c, \tilde{c}\rb\,
\exp \left[-S_{\msc{gh}}(b,\tilde{b},c,\tilde{c})\right] = \tau_2\left|\eta(\tau)\right|^4.
\label{Z gh}
\end{eqnarray}

\end{description}

~


\section*{Appendix C:~ Irreducible and Extended Characters and their Modular Completions}

\setcounter{equation}{0}
\def\theequation{C.\arabic{equation}}


In this appendix we summarize the definitions as well as 
useful formulas for the (extended) characters 
and their modular completions of the $\cN=2$ superconformal algebra with 
$\hc \left(\equiv \frac{c}{3} \right)= 1+ \frac{2}{k}$.
We focus only on  the $\tR$-sector\footnote
   {In this paper 
we shall use the convention of $\tR$-characters 
with the inverse sign compared to those of \cite{ES-NH,ES-BH,ES-C}, 
so that the Witten indices appear with the positive sign. 
(See \eqn{WI} below.)}, and when treating the extended characters, 
we assume $k= N/K$, $(N,K \in \bz_{>0}) $ (but, not assume $N$ and $K$ are co-prime).

~


\noindent
{\bf \underline{Continuous (non-BPS) Characters:}}
\begin{equation}
\ch {}{} (P,\mu;\tau,z) := q^{\frac{P^2+ \mu^2}{4k}} 
y^{\frac{\mu}{k}}  \,
\frac{\th_1(\tau,z)}{i \eta(\tau)^3},
\label{ch c}
\end{equation}
which is associated to the irrep. with the following conformal weight $h$ 
and $U(1)$-charge $Q$; 
\begin{equation}
h= \frac{P^2+ \mu^2}{4k}+\frac{\hc}{8}, ~~~ Q= \frac{\mu}{k} \pm \frac{1}{2},
~~(\mbox{doubly degenerated}) 
\end{equation}

The modular transformation formulas and the spectral flow property are 
summarized as
\begin{eqnarray}
&& 
\hspace{-1cm}
\ch{}{}\left(P,\mu ; - \frac{1}{\tau}, \frac{z}{\tau}\right)
= (-i) e^{i\pi \frac{\hc}{\tau}z^2}\, 
 \frac{1}{2k} \int_{-\infty}^{\infty} dP' \, 
\int_{-\infty}^{\infty} d\mu' \, 
e^{2\pi i \frac{P P' - \mu \mu'}{2k}}
\, \ch{}{} (P',\mu';\tau,z) .
\label{S ch}
\\
&& \hspace{-1cm}
\ch{}{}\left(P,\mu ; \tau+1, z \right)
= e^{2\pi i \frac{P^2+\mu^2}{4k}}\,
\ch{}{} \left(P, \mu ; \tau, z \right).
\label{T ch}
\\
&&
\hspace{-1cm}
\ch{}{} (P,\mu;\tau,z+r\tau+s) = (-1)^{r+s} e^{2\pi i \frac{\mu}{k}s}
q^{-\frac{\hc}{2}r^2} y^{-\hc r}\, \ch{}{}(P,\mu+2r;\tau,z),
~~~ (\any r,s \in \bz).
\label{sflow ch}
\end{eqnarray}

~

\noindent
{\bf \underline{Discrete Characters: } \cite{BFK,Dobrev}}
\begin{equation}
\chd (\la,n;\tau,z) := \frac{(yq^n)^{\frac{\la}{k}}}{1-yq^n}\,
 y^{\frac{2n}{k}}  q^{\frac{n^2}{k}}\,
\frac{\th_1(\tau,z)}{i \eta(\tau)^3},
\label{ch d}
\end{equation}
which is associated to the $n$-th spectral flow of 
discrete irrep. generated by the Ramond vacua\footnote
   {The unitarity requires $- \frac{\hc}{2} \leq Q \leq \frac{\hc}{2}$ for the Ramond 
vacua, which is equivalent with the condition : $-1 \leq \la \leq k+1$ \cite{BFK}.
The quantum number  $\la$ is identified with $2j-1$, where $j$ is the `isospin' 
of $SL(2,\br)$ in the $SL(2,\br)/U(1)$-coset \cite{DPL}. Thus, the unitarity range  
$-1 \leq \la \leq k+1$ corresponds to the `analogue of integrable representations'
$
0\leq j \leq \frac{k+2}{2}\equiv \frac{\kappa}{2},
$
where $\kappa$ denotes the level of bosonic $SL(2,\br)$-WZW. 
The range $0\leq \la \leq k ~ (\Leftrightarrow ~ \frac{1}{2} \leq j \leq \frac{k+1}{2} )$ that we adopt here 
is strictly narrower than this `unitarity range'.
This restriction has a clear origin in the discrete spectrum of  the SUSY $SL(2,\br)/U(1)$-coset
read off from the torus partition function. It is worth pointing out that this range 
is invariant under modular transformations given below. We also note that the missing `edge' points
$\la=-1,k+1$ correspond to the `graviton representation' and its spectral flows, which obey 
different character formulas (see \cite{Dobrev}). 
This type restriction of spectrum has been already discussed in 
\cite{GK,MO,ES-BH}. };
\begin{equation}
h= \frac{\hc}{8}, ~~~ 
Q = \frac{\la}{k} - \frac{1}{2}, ~~(0\leq \la \leq k) 
\end{equation}


The modular transformation formulas are given as \cite{ES-L} 
\begin{eqnarray}
&& 
\hspace{-1cm}
\chd \left(\la,n ; - \frac{1}{\tau}, \frac{z}{\tau}\right)
= e^{i\pi \frac{\hc}{\tau}z^2}\,\left[
 \frac{1}{k} \int_0^k d\la'  \,\sum_{n' \in \bz}\,
e^{2\pi i \frac{\la \la' - (\la+2n)(\la'+2n')}{2k}}
\, \chd (\la',n';\tau,z) \right.
\nn
&& \hspace{1.5cm}
\left. -\frac{i}{2k} \, \int_{-\infty}^{\infty} d\mu'
\, e^{-2\pi i \frac{(\la+2n) \mu'}{2k}}\,\int_{\br+i0} dP'\,
 \frac{e^{-2\pi \frac{\la P'}{2k}}}{1-e^{-\pi (P'+ i\mu')}}
\, \ch{}{} (P',\mu';\tau,z) \right],
\label{S ch d}
\\
&& \hspace{-1cm}
\chd \left(\la,n ; \tau+1, z \right)
= e^{2\pi i \frac{n}{k} \left(\la+ n \right)}\,
\chd \left(\la,n ; \tau, z \right).
\label{T ch d}
\end{eqnarray}
The spectral flow property is written as 
\begin{equation}
\chd (\la,n;\tau,z+r\tau+s) = (-1)^{r+s} e^{2\pi i \frac{\la+2n}{k}s}
q^{-\frac{\hc}{2}r^2} y^{-\hc r}\, \chd(\la,n+r;\tau,z),
~~~ (\any r,s \in \bz).
\label{sflow chd}
\end{equation}

~


\noindent
{\bf \underline{Extended Continuous (non-BPS) Characters \cite{ES-L,ES-BH}:}}
\begin{eqnarray}
\chic (p,m;\tau,z) &:=  & \sum_{n\in N\bz}\, 
(-1)^n q^{\frac{\hc}{2}n^2} y^{\hc n} \, \ch{}{}\left(\frac{p}{K}, \frac{m}{K}; \tau, z+n\tau\right)
\nn
&=& 
q^{\frac{p^2}{4NK}} \Th{m}{NK}\left(\tau,\frac{2z}{N}\right)\,
\frac{\th_1(\tau,z)}{i \eta(\tau)^3}.
\label{chic}
\end{eqnarray}
This corresponds to the spectral flow sum of the non-degenerate representation with
$h= \frac{p^2+m^2}{4NK} + \frac{\hc}{8}$, 
$Q = \frac{m}{N}\pm \frac{1}{2}$~($p\geq 0$, $m\in \bz_{2NK}$),
whose flow momenta are taken to be $n\in N \bz$.
The modular and spectral flow properties are simply written as 
\begin{eqnarray}
&& 
\hspace{-1cm}
\chic\left(p,m ; - \frac{1}{\tau}, \frac{z}{\tau}\right)
= (-i) e^{i\pi \frac{\hc}{\tau}z^2}\, 
 \frac{1}{2NK} \int_{-\infty}^{\infty} dp' \, 
\sum_{m'\in\bz_{2NK}}\, 
e^{2\pi i \frac{p p' - m m'}{2NK}}
\, \chic (p',m';\tau,z) .
\label{S chic}
\\
&& \hspace{-1cm}
\chic\left(p,m ; \tau+1, z \right)
= e^{2\pi i \frac{p^2+m^2}{4NK}}\,
\chic \left(p, m ; \tau, z \right),
\label{T chic}
\\
&& \hspace{-1cm}
\chic (p,m;\tau,z+r\tau+s) = (-1)^{r+s} e^{2\pi i \frac{m}{N}s}
q^{-\frac{\hc}{2}r^2} y^{-\hc r}\, \chic(p,m+2Kr;\tau,z),
~~~ (\any r,s \in \bz).
\nn
&&
\label{sflow chic}
\end{eqnarray}

~


\noindent
{\bf \underline{Extended Discrete (BPS) Characters \cite{ES-L,ES-BH,ES-C}:}}
\begin{eqnarray}
\chid (v,a;\tau,z) &:= & 
 \sum_{n\in N\bz}\, 
(-1)^n q^{\frac{\hc}{2}n^2} y^{\hc n} \, \chd\left(\frac{v}{K}, a ; \tau, z+n\tau\right)
\nn
&=&  \sum_{n\in \bz}\,\chd\left(\frac{v}{K}, a+N n ; \tau, z\right)
\nn
&=& \sum_{n\in\bz}\, \frac{(yq^{N n+ a})^{\frac{v}{N}}}
{1-yq^{Nn+a}} \, y^{2K\left(n+\frac{a}{N}\right)} q^{NK \left(n+\frac{a}{N}\right)^2}
\, \frac{\th_1(\tau,z)}{i \eta(\tau)^3}.
\label{chid}
\end{eqnarray}
This again corresponds to the  sum of the 
Ramond vacuum representation with $h= \frac{\hc}{8}$, $Q= \frac{v}{N}-\frac{1}{2}$
~($v=0,1,\ldots , N-1$) over spectral flow
 with flow momentum $m$  taken to be mod.$N$, as 
$m= a +N\bz$ ~ ($a\in \bz_N$).

The modular transformation formula 
can be expressed as \cite{ES-L,ES-BH,ES-C};
\begin{eqnarray}
&& 
\hspace{-1cm}
\chid \left(v,a ; - \frac{1}{\tau}, \frac{z}{\tau}\right)
= e^{i\pi \frac{\hc}{\tau}z^2}\,\left[
 \sum_{v=0}^{N-1} \,\sum_{a\in \bz_N}\,
\frac{1}{N} \, e^{2\pi i \frac{vv' - (v+2Ka)(v'+2Ka')}{2NK}}
\, \chid  (v',a';\tau,z) \right.
\nn
&& \hspace{1.5cm}
\left. -\frac{i}{2NK} \sum_{m' \in \bz_{2NK}} \, e^{-2\pi i \frac{(v+2Ka) m'}{2NK}}\,
\int_{\br+i0} dp'\, \frac{e^{-2\pi \frac{vp'}{2NK}}}{1-e^{-\pi \frac{p'+im'}{K}}}
\, \chic (p',m';\tau,z)
\right],
\label{S chid}
\\
&& 
\hspace{-1cm}
\chid \left(v,a ; \tau+1, z \right)
= e^{2\pi i \frac{a}{N} \left(v+ K a \right)}\,
\chid \left(v,a ; \tau, z \right),
\label{T chid} 
\end{eqnarray}
%

The spectral flow property is also expressed as 
\begin{equation}
\chid (v,a;\tau,z+r\tau+s) = (-1)^{r+s} e^{2\pi i \frac{v+2Ka}{N}s}
q^{-\frac{\hc}{2}r^2} y^{-\hc r}\, \chid(v,a+r;\tau,z),
~~~ (\any r,s \in \bz),
\label{sflow chid}
\end{equation}

~

\noindent
{\bf \underline{Modular Completion of the Irreducible Discrete Character \eqn{ch d}:}}
\begin{eqnarray}
\hspace{-5mm}
\hchd(\la,n;\tau,z) &:= & 
\chd(\la,n;\tau,z) 
\nn
&& 
- \frac{1}{2} \sum_{\nu \in \la + k \bz}\, \sgn(\nu +0)\, 
\erfc \left(\sqrt{\frac{\pi \tau_2}{k}} \left|\nu \right|\right)\, 
q^{\frac{n^2}{k}+ \frac{n}{k}\nu}\,
y^{\frac{1}{k} (\nu +2n)}\, \frac{\th_1(\tau,z)}{i \eta(\tau)^3} 
\nn
&= & \frac{\th_1(\tau,z)}{i\eta(\tau)^3} \, y^{\frac{2n}{k}}q^{\frac{n^2}{k}}
\, \left\lb 
\frac{(yq^n)^{\frac{\la}{k}}}{1-yq^n} 
-\frac{1}{2} \sum_{\nu \in\la + k\bz}\, \sgn(\nu +0)\, 
\erfc \left(\sqrt{\frac{\pi \tau_2}{k}} \left|\nu \right| \right)\, 
(y q^n)^{\frac{\nu}{k}}
\right\rb
\nn
&=  &  \frac{ \th_1(\tau,z)}{ 2\pi \eta(\tau)^3}\,
\frac{y^{\frac{2n}{k}} q^{\frac{n^2}{k}}}{1-y q^{n}}
\sum_{\nu \in \la + k\bz}\, \left\{ \int_{\br + i(k-0)} dp\, -
\int_{\br-i0} dp \, \left(y q^{n} \right) 
\right\}
\,
\frac{ e^{- \pi \tau_2 \frac{p^2+\nu^2}{k}} \left(y q^{n}\right)^{\frac{\nu }{k}}}{p-i\nu} ,
\nn
&& 
\hspace{7cm}
(0\leq \la \leq k, ~~n\in \bz),
\label{hchd}
\end{eqnarray}
where $\erfc(*)$ denotes the error-function defined by
\begin{equation}
\erfc(x):= \frac{2}{\sqrt{\pi}} \int_x^{\infty} e^{-t^2} \, dt
\left( \equiv 1- \erf(x) \right).
\label{erfc}
\end{equation}
The equality in the last line of \eqn{hchd} is derived from the integral formula;
\begin{equation}
\int_{\br\mp i0} dp\, \frac{e^{-\al p^2}}{p-i\nu} = i\pi e^{\al \nu^2} \sgn(\nu \pm 0) 
\erfc(\sqrt{\al}|\nu|), 
\hspace{1cm}(\nu \in \br, ~ \al >0),
\label{formula erfc}
\end{equation}
and by using a simple contour deformation technique.

Note that $\hchd(\la,n;\tau,z) $ is 
non-holomorphic due to the explicit dependence on 
$\tau_2 \equiv \Im \tau$.
It is crucial that the modular completion $\hchd$ has nice modular properties. 
Especially, one can prove that the S-transformation formula gets considerably simplified\footnote
   {Probably, the easiest way to prove it would  be given by regarding $\hchd(\la,n;\tau,z) $ as 
the `continuum limit' of $\hchid(v,a;\tau,z)$. (See \eqn{rel hchid hchd}.) 
The modular property of $\hchid(v,a;\tau,z)$ can be straightforwardly read off from the expansion formula of elliptic genus \eqn{tcZ bhchi}, and  
we eventually arrive at the wanted formula \eqn{S hchd}.
};
\begin{eqnarray}
&& 
\hspace{-1cm}
\hchd \left(\la,n ; - \frac{1}{\tau}, \frac{z}{\tau}\right)
= e^{i\pi \frac{\hc}{\tau}z^2}\,
 \frac{1}{k} \int_0^k d\la'  \,\sum_{n' \in \bz}\,
e^{2\pi i \frac{\la \la' - (\la+2n)(\la'+2n')}{2k}}
\, \hchd (\la',n';\tau,z).
\label{S hchd}
\end{eqnarray}
It is easy to see that the T-transformation and spectral flow property 
are preserved by taking the completion;
\begin{eqnarray}
&& \hchd \left(\la,n ; \tau+1, z \right)
= e^{2\pi i \frac{n}{k} \left(\la+ n \right)}\,
\hchd \left(\la,n ; \tau, z \right),
\label{T hchd}
\\
&& \hchd (\la,n;\tau,z+r\tau+s) = (-1)^{r+s} e^{2\pi i \frac{\la+2n}{k}s}
q^{-\frac{\hc}{2}r^2} y^{-\hc r}\, \hchd(\la,n+r;\tau,z),
\nn
&& \hspace{10cm} (\any r,s \in \bz).
\label{sflow hchd}
\end{eqnarray}

~


\noindent
{\bf \underline{Modular Completion of the Extended Discrete Characters:}}

The modular completion of the discrete character $\chid$ is defined as 
the spectral flow sum of $\hchd$ \eqn{hchd} in the similar manner to \eqn{chid};
\begin{eqnarray}
\hspace{-1cm}
\hchid (v,a;\tau,z) 
&:= & \sum_{n\in N\bz} \, (-1)^n q^{\frac{\hc}{2}n^2} y^{\hc n}\,
\hchd\left(\frac{v}{K}, a;\tau,z+n\tau\right)
\nn
&=&
 \sum_{m\in\bz}\, \hchd\left(\frac{v}{K}, a + Nm ;\tau,z\right)
\nn
&=  &  \chid (v,a;\tau,z) - \frac{1}{2} \sum_{j\in \bz_{2K}}\,
R_{v+Nj, NK}(\tau) \Th{v+Nj+2Ka}{NK}\left(\tau, \frac{2z}{N}\right)\,
\frac{\th_1(\tau,z)}{i \eta(\tau)^3},
\nn
&=& 
\frac{\th_1(\tau,z)}{ 2\pi \eta(\tau)^3}\,
\sum_{\stackrel{n\in a + N\bz}{s\in v + N\bz}}\, 
\frac{y^{\frac{2Kn}{N}} q^{\frac{K n^2}{N}}}{1-y q^{n}}
\left\{ \int_{\br + i(N-0)} dp\, -
\int_{\br-i0} dp \, \left(y q^{n} \right) 
\right\}
\,
\frac{ e^{- \pi \tau_2 \frac{p^2+s^2}{NK}} 
\left(y q^{n}\right)^{\frac{s}{N}}}{p-is} ,
\nn
&&
\label{hchid}
\end{eqnarray}
where we set 
\begin{eqnarray}
R_{m,k}(\tau) &:=& \sum_{\nu \in m+2k\bz}\, \sgn(\nu + 0) 
\erfc\left(\sqrt{\frac{\pi \tau_2}{k}} \left|\nu\right|\right)\, q^{- \frac{\nu^2}{4k}}
\nn
&=& \frac{1}{i\pi}\, \sum_{\nu \in m+2k\bz}\,
\int_{\br- i0} dp \, \frac{e^{-\pi \tau_2 \frac{p^2+\nu^2}{k}} }{p-i\nu}\,
q^{- \frac{\nu^2}{4k}}.
\label{Rmk}
\end{eqnarray}
%

Conversely the irreducible modular completion \eqn{hchd}
is reconstructed from the extended one \eqn{hchid} by taking the `continuum limit' ;
\begin{equation}
\lim_{\stackrel{N\,\rightarrow\, \infty}{k\equiv N/K }\, \msc{fixed}}\, 
\hchid\left(v,a;\tau,z\right) = \hchd \left(\la \equiv \frac{v}{K}, a ;\tau,z\right).
\label{rel hchid hchd} 
\end{equation}

In the main text of this paper we also use an alternative notation
\begin{eqnarray}
&& 
\bhchi(v,m;\tau,z) \equiv \hchid(v,a;\tau,z),  ~ ~  \mbox{with} ~ 
m\equiv  v+2Ka \in \bz_{2NK}, ~~~ v=0, 1 \ldots, N-1,
\nn
&& \bhchi(v,m;\tau,z) \equiv 0, ~~ \mbox{if} ~ m-v \not\in 2K \bz.
\label{bhchi def}
\end{eqnarray}

The modular transformation formulas for \eqn{hchid} are written as 
\begin{eqnarray}
&& \hspace{-1cm}
\hchid \left(v,a ; - \frac{1}{\tau}, \frac{z}{\tau}\right)
= e^{i\pi \frac{\hc}{\tau}z^2}\, \sum_{v'=0}^{N-1} \,\sum_{a'\in \bz_N}\,
\frac{1}{N} \, e^{2\pi i \frac{vv' - (v+2Ka)(v'+2Ka')}{2NK}}
\, \hchid (v',a';\tau,z),
\label{S hchid}
\\
&& \hspace{-1cm}
\hchid \left(v,a ; \tau+1, z \right)
= e^{2\pi i \frac{a}{N} \left(v+ K a \right)}\,
\hchid \left(v,a ; \tau, z \right).
\label{T hchid}
\end{eqnarray}
Also the spectral flow property is preserved by taking the completion;
\begin{equation}
\hchid (v,a;\tau,z+r\tau+s) = (-1)^{r+s} e^{2\pi i \frac{v+2Ka}{N}s}
q^{-\frac{\hc}{2}r^2} y^{-\hc r}\, \hchid(v,a+r;\tau,z),
~~~ (\any r,s \in \bz).
\label{sflow hchid}
\end{equation}

It may be useful to note the following identity of
$R_{m,k}(\tau)$ \cite{Zwegers}, which relates the 
S-transformation of $R_{m,k}(\tau)$ to the Mordell integral
 \cite{Mordell,Watson};
\begin{eqnarray}
&&
R_{m,k}(\tau)+{i\over \sqrt{-i\tau}}{1\over \sqrt{2k}}\sum_{\ell \in \bz_{2k}}
\, e^{-{i \pi m\ell\over k}}R_{\ell,k}\left(- {1\over \tau}\right)
=2ie^{-{i \pi  m^2\tau\over 2k}}\int_{\br-it} dp\, 
{e^{2\pi i k\tau p^2-2\pi m \tau p}\over 1-e^{2\pi p}}, 
\nn 
&& \hspace{11cm}  (0<\any t<1).
\label{id R Mordell}
\end{eqnarray}

~

\noindent
{\bf \underline{Witten Index :}}
\begin{eqnarray}
&& \lim_{z\,\rightarrow\, 0}\, \chd (\la,n;\tau,z) = 
\lim_{z\,\rightarrow\, 0}\, \hchd (\la,n;\tau,z) = \delta_{n,0},
\nn
&& \lim_{z\,\rightarrow\, 0}\, \chid (v,a;\tau,z) = 
\lim_{z\,\rightarrow\, 0}\, \hchid (v,a;\tau,z) =  \delta_{a,0}^{(N)}
\equiv \left\{
\begin{array}{ll}
1 & ~~~ a\equiv 0~ (\mod N) \\
0 & ~~~ \mbox{otherwise}.
\end{array}
\right.
\label{WI}
\end{eqnarray}

~


\section*{Appendix D:~ Detailed Calculations for the `Character Decomposition'}

\setcounter{equation}{0}
\def\theequation{D.\arabic{equation}}


In this appendix we present a detailed calculation to derive 
the discrete parts of partition functions. It is almost parallel to  those given in
\cite{ES-NH}. We shall only work with 
\eqn{tZ beta}, and other formulas are reproduced from it. 
We again assume $k=N/K$, $N=ML$ 
with some positive integers $N,K,M,L$.

Let us start with rewriting \eqn{tZ beta} in 
the form of `Fourier transformation' (recall \eqn{Fourier trsf 1});
\begin{eqnarray}
&& 
\tZ^{(M)}_{\reg}(\tau,z\,| \al, \beta ;\ep) 
= 
\frac{k}{M} \, e^{- \frac{2\pi}{\tau_2} \frac{k+4}{k} z_2^2}\,
\sum_{w, m \in \bz} \,
e^{2\pi i \frac{1}{M}(w \beta- m\al)}
\,  
\nn
&& \hspace{2.5cm} \times
\int_{\Sigma(z,\ep)} \frac{d^2 u}{\tau_2} \, 
\left|
\frac{\th_1\left(\tau,u+\frac{k+2}{k}z\right)}{\th_1\left(\tau, u+\frac{2}{k}z\right)}
\right|^2 \,
e^{-4\pi \frac{u_2z_2}{\tau_2}} \, e^{-\frac{\pi k}{\tau_2} \left|u+ 
\frac{w\tau+m}{M}\right|^2}.
\label{tZ 2}
\end{eqnarray}
After making a small change of integration variables, 
we obtain
\begin{eqnarray}
\hspace{-5mm}
\tZ^{(M)}_{\reg}(\tau, z\,|\, \al,\beta;\ep) &=& 
\frac{k}{M} e^{-\frac{2\pi}{\tau_2} z_2^2}\,
\sum_{w,m \in \bsz}\, e^{2\pi i \frac{1}{M}(w \beta- m \al)}\,
\int_\ep^{1-\ep} ds_1  \, \int_0^1 ds_2\, 
\left| \frac{\th_1\left(\tau, -s_1\tau-s_2+ z \right)}
{\th_1\left(\tau, -s_1\tau-s_2 \right)} \right|^2 
\nn
&& \hspace{3cm}
\times  e^{4\pi s_1 z_2}
\, e^{-\frac{\pi k }{\tau_2} \left|\left(s_1+ \frac{w}{M} \right)\tau 
+ \left(s_2+ \frac{m}{M} \right) + \frac{2}{k}z\right|^2}.
\nonumber
\end{eqnarray}
By dualizing the temporal winding number $m$ into the KK momentum $n$
by means of the Poisson resummation formula, 
we can further rewrite it as 
\begin{eqnarray}
\hspace{-5mm}
\tZ^{(M)}_{\reg}(\tau,z\,| \al, \beta ;\ep) &=& 
\sqrt{k\tau_2} 
e^{-\frac{2\pi}{\tau_2} z_2^2 }\,
\sum_{w,n \in \bsz}\, e^{2\pi i \frac{w \beta}{M}}\,
\int_{\ep}^{1-\ep}ds_1 \, \int_{0}^{1} ds_2\, 
\left| \frac{\th_1\left(\tau, -s_1\tau-s_2+ z\right)}
{\th_1\left(\tau, -s_1\tau-s_2 \right)} \right|^2 
\nn
&& \hspace{1cm}
\times  e^{4\pi s_1 z_2}
\, e^{-\pi \tau_2 \left\{ \frac{n^2}{k} + k \left(s_1+ \frac{w}{M}  
+ \frac{2z_2}{k\tau_2}\right)^2\right\}
+ 2\pi i n \left\{\left(s_1+\frac{w}{M} \right)\tau_1 
+ s_2 + \frac{2z_1}{k} \right\}}
\nn
&=& 
\sqrt{k\tau_2} 
e^{-2\pi \frac{\hc}{\tau_2}z_2^2} \,
\sum_{w,n \in \bsz}\, e^{2\pi i \frac{w \beta}{M}}\,
\int_{\ep}^{1-\ep}ds_1 \, \int_{0}^{1} ds_2\, 
\left| \frac{\th_1\left(\tau, -s_1\tau-s_2+ z\right)}
{\th_1\left(\tau, -s_1\tau-s_2 \right)} \right|^2 
\nn
&& \hspace{0.5cm}
\times  
\, e^{-\pi \tau_2 \left\{ \frac{1}{k}(Mn+\al)^2 
+ k \left(s_1+ \frac{w}{M}  \right)^2\right\}
+ 2\pi i (Mn+\al) \left\{\left(s_1+ \frac{w}{M} \right)\tau_1 
+ s_2 \right\}} \,
y^{\frac{w}{M}+\frac{Mn+\al}{k}} \bar{y}^{\frac{w}{M}-\frac{Mn+\al}{k}}.
\nn
&&
\label{tZ 4}
\end{eqnarray}
Substituting  the identity\footnote
  {See {\em e.g.} \cite{ES-NH}.}; 
($u\equiv s_1\tau+s_2$, $0< s_1 < 1$);
\begin{eqnarray}
&& \left| \frac{\th_1\left(\tau, -s_1\tau-s_2+ z\right)}
{\th_1\left(\tau, -s_1\tau-s_2 \right)} \right|^2 = 
\left|\frac{\th_1(\tau,z)}{i \eta(\tau)^3}\right|^2\,
\sum_{\ell, \tell \in \bz} \, \frac{yq^{\ell}} 
{1-yq^{\ell}}\cdot \left[{y q^{\tell}} \over{1- y q^{\tell}}\right]^*
\nn
&& \hspace{4cm}
\times 
e^{-2\pi i (s_1\tau_1+s_2) (\ell-\tell) + 2\pi s_1 \tau_2 (\ell+\tell)}.
\label{exp th1/th1}
\end{eqnarray}
into \eqn{tZ 4}, one can easily  
integrate $s_2$ out, which just yields the constraint 
\begin{equation}
Mn+ \al = \ell-\tell.
\label{cond n l}
\end{equation}

We next evaluate the $s_1$-integral. 
Picking up relevant terms, we obtain
\begin{eqnarray}
&& e^{-\pi \tau_2 \frac{N}{K} s_1^2 
-2\pi s_1 \left\{\tau_2 \frac{L}{K} w - i\tau_1 (M n+\al) 
+ i \tau_1 (\ell-\tell)-\tau_2(\ell+\tell) \right\}
} 
= e^{-\pi \tau_2 \frac{N}{K} s_1^2 - 2\pi s_1 \tau_2 \frac{v}{K}},
\end{eqnarray}
where we set 
\begin{equation}
v:= Lw- K(\ell+\tell),
\label{v def}
\end{equation}
and used the condition \eqn{cond n l}. 
Utilizing a Gaussian integral, we  obtain
\begin{eqnarray}
\int_{\ep}^{1-\ep} ds_1 
\, e^{-\pi \tau_2 \frac{N}{K} s_1^2 -2\pi s_1\tau_2 \frac{v}{K}}
&=& \sqrt{\frac{\tau_2}{NK}} \,  \int_{\ep}^{1-\ep} ds_1\, \int_{\br-i0} dp\,
e^{-\frac{\pi}{NK}\tau_2 p^2 - 2\pi i \tau_2 \frac{s_1}{K}(p-iv)}
\nonumber \\
&=& \sqrt{\frac{K}{N\tau_2}} \frac{1}{2\pi i} \,
\int_{\br-i0} dp\, 
\frac{e^{-\frac{\pi}{NK} \tau_2 p^2  }} {p-iv}
\, \left\{ 
e^{-\vep(v+ip)}
- e^{\vep(v+ip)}
e^{-2\pi i \tau_2 \frac{1}{K} (p-iv)}
\right\},
\nn
&&
\label{s1 int} 
\end{eqnarray}
where we set $\vep \equiv 2\pi \frac{\tau_2}{K} \ep \, (>0)$.

Collecting remaining exponents of $q$ and $y$, we further obtain the  factor;
$$
e^{- \pi \tau_2 \frac{v^2}{NK}}\, q^{\frac{1}{N}\left(K\ell^2 + \ell v\right)} 
\bar{q}^{\frac{1}{N}\left(K\tell^2 + \tell v\right)} 
\, y^{\frac{2K}{N}\left(\ell + \frac{v}{2K}\right)}
\bar{y}^{\frac{2K}{N}\left(\tell + \frac{v}{2K}\right)}.
$$

Combining all the pieces we finally obtain 
\begin{eqnarray}
&& \hspace{-1.5cm}
\tZ^{(M)}_{\reg}(\tau,z\,|\, \al,\beta \, ; \ep) = 
e^{-2\pi \frac{\hc}{\tau_2}z_2^2} \,
\left|\frac{\th_1(\tau,z)}{\eta(\tau)^3}\right|^2 \,
\sum_{\stackrel{\ell,\tell\in \bz}{\ell-\tell \in \al+ M\bz}}\,
\sum_{\stackrel{v\in \bz}{v+K(\ell+\tell) \in L\bz}}\,
\nn
&&
\hspace{1.5cm} 
\times 
\frac{1}{2\pi i} \, \left[
\int_{\br-i0} dp \, e^{-\vep (v+ip)} yq^{\ell}\, \left[yq^{\tell}\right]^*
- \int_{\br+i(N-0)} dp \,  e^{\vep (v+ip)}
\right]
\nn
&& \hspace{1.5cm}
\times 
e^{2\pi i \frac{\beta}{N} \left\{v+ K(\ell+\tell) \right\}}
\,
\frac{e^{-\pi \tau_2 \frac{p^2+v^2}{NK}}} {p-iv}
\, \frac{(yq^{\ell})^{\frac{v}{N}}}{1-yq^{\ell}} \,
\left[
\frac{(yq^{\tell})^{\frac{v}{N}}}{1-yq^{\tell}}
\right]^*\,
y^{\frac{2K}{N}\ell} q^{\frac{K}{N}\ell^2} \, 
\left[y^{\frac{2K}{N}\tell} q^{\frac{K}{N}\tell^2}\right]^*.
\label{tZ final}
\end{eqnarray}
Here the factor
$$
yq^{\ell} \cdot \left[y q^{\tell}\right]^* \cdot e^{-2\pi i \tau_2 \frac{1}{K}(p-iv)}
$$
was absorbed into 
the change of 
variables $p=: p'-iN$, $v=:  v'- N$ with
\begin{eqnarray*}
&& \frac{p^2+v^2}{NK} = \frac{p^{'2}+v^{'2}}{NK} - \frac{2i}{K}(p'-iv'), \hspace{1cm}
p-iv = p'-iv', 
\\
&& \left(yq^{\ell}\right)^{\frac{v}{N}} \, 
\left[\left( y q^{\tell}\right)^{\frac{v}{N}}\right]^* 
=  \left(yq^{\ell}\right)^{\frac{v'}{N}-1} \,
\left[ \left( y q^{\tell}\right)^{\frac{v'}{N}-1}\right]^* .
\end{eqnarray*}


At this stage, one can successfully extract a sesquilinear form of the modular completion
$\hchid$  from \eqn{tZ final} as the discrete part in a similar manner to \eqn{part fn A decomp 0},
that is, 
$$
\tZ^{(M)}_{\reg}(\tau,z\,|\, \al,\beta;\ep) = \tZ_{\dis}^{(M)} (\tau,z\,|\, \al,\beta) 
+ [\mbox{sesquilinear form of $\chic(\tau,z)$}].
$$ 
Performing a `completion of the square' 
by using the last line of \eqn{hchid},  
one can achieve 
\begin{equation}
\tZ^{(M)}_{\dis} (\tau, z\,|\, \al,\beta) =  
e^{- 2\pi \frac{\hc}{\tau_2}z_2^2} \, 
\sum_{(v,a, \ta) \in \cR (M,\al)}
\, e^{2\pi i \frac{\beta}{N} \left\{v+K(a+\ta)  \right\}} \,
\hchid (v,a;\tau,z) \left[\hchid (v,  \ta ;\tau,z)\right]^*,
\label{tZ dis al beta}
\end{equation}
where the range of summation $\cR(M,\al)$ is given in \eqn{R M al}, namely,
\begin{eqnarray*}
\cR(M,\al) & = & \left\{
(v, a, \ta) \in \bz \times \bz_{N} \times \bz_{N}~;~
0\leq v \leq N-1, \right. 
\nn
&& \hspace{2cm} \left. 
a-\ta \equiv \al~ (\mod M), ~   v+K(a+\ta) \in L\bz
\right\}.
\end{eqnarray*}
This is the desired formula \eqn{tZdis bhchi}.

~


\newpage
\begin{thebibliography}{99}




\bibitem{ES-NH}
  T.~Eguchi and Y.~Sugawara,
  JHEP {\bf 1103}, 107 (2011)
  [arXiv:1012.5721 [hep-th]].




\bibitem{2DBH}
E. Witten, 
Phys. Rev. {\bf D44} (1991) 314;
G. Mandal, A. Sengupta and S. Wadia, 
Mod. Phys. Lett. {\bf A6} (1991) 1685;
I. Bars and D. Nemeschansky, 
Nucl. Phys. {\bf B348} (1991) 89;
S. Elizur, A. Forge and E. Rabinovici, 
Nucl. Phys. {\bf B359} (1991) 581;
R.~Dijkgraaf, H.~Verlinde and E.~Verlinde,
Nucl.\ Phys.\ B {\bf 371}, 269 (1992).




\bibitem{Troost}
  J.~Troost,
  JHEP {\bf 1006}, 104 (2010)
  [arXiv:1004.3649 [hep-th]].

\bibitem{AshokTroost}
  S.~K.~Ashok, J.~Troost,
  JHEP {\bf 1103}, 067 (2011).
  [arXiv:1101.1059 [hep-th]].



\bibitem{HPT}
A.~Hanany, N.~Prezas and J.~Troost,
JHEP {\bf 0204}, 014 (2002)
[arXiv:hep-th/0202129].



\bibitem{IPT}
D.~Israel, A.~Pakman and J.~Troost,
  JHEP {\bf 0404}, 043 (2004)
  [arXiv:hep-th/0402085].



\bibitem{ES-BH}
  T.~Eguchi and Y.~Sugawara,
  JHEP {\bf 0405}, 014 (2004)
  [arXiv:hep-th/0403193].


\bibitem{IKPT}
  D.~Israel, C.~Kounnas, A.~Pakman and J.~Troost,
  JHEP {\bf 0406}, 033 (2004)
  [arXiv:hep-th/0403237].



\bibitem{ES-C}
  T.~Eguchi and Y.~Sugawara,
  JHEP {\bf 0501}, 027 (2005)
  [arXiv:hep-th/0411041].



\bibitem{ES-L}
T.~Eguchi and Y.~Sugawara,
JHEP {\bf 0401}, 025 (2004)
[arXiv:hep-th/0311141].





\bibitem{Mordell}
L.J. Mordell, Quarterly Journal of Math. {\bf  68}, 329 (1920).



\bibitem{Watson}
G.N. Watson, J. London Math. Soc. {\bf 11}, 55 (1936).




\bibitem{ET}
T.~Eguchi and A.~Taormina,
Phys.\ Lett.\ B {\bf 200}, 315 (1988);
Phys.\ Lett.\ B {\bf 210}, 125 (1988).






\bibitem{EZ}
M. Eichler and D. Zagier, 
{\it``The Theory of Jacobi Forms,''}
(Birkh\"{a}user, 1985).


\bibitem{Zwegers}
S. Zwegers, PhD thesis, 
``Mock Theta functions'', Utrecht University, 2002.


\bibitem{BOno}
K. Bringman and K. Ono,
``Lifting cusp forms to Maass forms with an application to partitions'',
Proceedings of the National Academy of Sciences of the United States of America
{\bf 104} (10) 3725-3731.




\bibitem{EH}
  T.~Eguchi and K.~Hikami,
  J.\ Phys.\ A  {\bf 42}, 304010 (2009)
  [arXiv:0812.1151 [math-ph]];
  T.~Eguchi and K.~Hikami,
  arXiv:0904.0911 [math-ph].
  
  


\bibitem{ZF}
 V.~A.~Fateev, A.~B.~Zamolodchikov,
  Sov.\ Phys.\ JETP {\bf 62}, 215-225 (1985).


\bibitem{GQ}
D.~Gepner and Z.~A.~Qiu,
Nucl.\ Phys.\ B {\bf 285}, 423 (1987).



\bibitem{Tduality-2DBH}
  A.~Giveon, M.~Porrati, E.~Rabinovici,
  Phys.\ Rept.\  {\bf 244}, 77-202 (1994).
  [hep-th/9401139], and references therein.





\bibitem{KS}
Y.~Kazama and H.~Suzuki,
Nucl.\ Phys.\ B {\bf 321}, 232 (1989).







\bibitem{KarS}
D.~Karabali and H.~J.~Schnitzer,
Nucl.\ Phys.\ B {\bf 329}, 649 (1990).


\bibitem{GawK}
K.~Gawedzki and A.~Kupiainen,
Nucl.\ Phys.\ B {\bf 320}, 625 (1989).


\bibitem{Schnitzer}
  H.~J.~Schnitzer,
  Nucl.\ Phys.\  B {\bf 324}, 412 (1989).






\bibitem{MMS}
  J.~M.~Maldacena, G.~W.~Moore and N.~Seiberg,
  JHEP {\bf 0107}, 046 (2001)
  [arXiv:hep-th/0105038].





\bibitem{ES-nh-P}
  T.~Eguchi and Y.~Sugawara,
  arXiv:1407.7721 [hep-th].




\bibitem{Witten-E}
E.~Witten,
Commun.\ Math.\ Phys.\  {\bf 109}, 525 (1987).



\bibitem{Appell}
A. ~Polishchuk,
arXiv:math.AG/9810084;
A.~M.~Semikhatov, A.~Taormina and I.~Y.~Tipunin,
arXiv:math.qa/0311314.




\bibitem{Gaw}
K.~Gawedzki,
arXiv:hep-th/9110076.




\bibitem{BFK}
W.~Boucher, D.~Friedan and A.~Kent,
Phys.\ Lett.\ B {\bf 172}, 316 (1986),


\bibitem{Dobrev}
V.~K.~Dobrev,
Phys.\ Lett.\ B {\bf 186}, 43 (1987);
E.~Kiritsis,
Int.\ J.\ Mod.\ Phys.\ A {\bf 3}, 1871 (1988).








\bibitem{DPL}
L.~J.~Dixon, M.~E.~Peskin and J.~Lykken,
Nucl.\ Phys.\ B {\bf 325}, 329 (1989).






\bibitem{GK}
A.~Giveon and D.~Kutasov,
JHEP {\bf 9910}, 034 (1999)
[arXiv:hep-th/9909110];
A.~Giveon and D.~Kutasov,
JHEP {\bf 0001}, 023 (2000)
[arXiv:hep-th/9911039].

\bibitem{MO}
J.~M.~Maldacena and H.~Ooguri,
J.\ Math.\ Phys.\  {\bf 42} (2001) 2929, hep-th/0001053.




\end{thebibliography}
\end{document}